\documentclass{tlp}

\usepackage[dvips]{epsfig} 

\newcommand{\lprolog}{$\lambda$Prolog}

\newcommand{\captionx}[2]{\def\figurename{#1}\caption{#2}}

\newcommand{\cd}[1]{\texttt{#1}}
\newcommand{\lamb}{{\tt \char'134}}

\def\mud#1{\hfil $\displaystyle{\mathstrut #1}$\hfil}
\def\rig#1{\hfil $\displaystyle{#1}$}
\def\vcalignhbox#1{\hbox{\valign{\vfil\hbox{##}\vfil\cr #1}}}
\def\vcalign#1{$$\vcalignhbox{#1}$$}

\newbox\tempa
\newbox\tempb
\newdimen\tempc
\newbox\tempd

\def\ruleanhelp#1#2#3{\setbox\tempa=%
                           \hbox{$\displaystyle{\mathstrut #2}$}%
		        \setbox\tempb=\vbox{\halign{##\cr
	\mud{#1}\cr
	\noalign{\vskip\the\lineskip}%
	\noalign{\hrule height 0pt}%
	\rig{\vbox to 0pt{\vss\hbox to 0pt{${\; #3}$\hss}\vss}}\cr
	\noalign{\hrule}%
	\noalign{\vskip\the\lineskip}%
	\mud{\copy\tempa}\cr}}%
		      \tempc=\wd\tempb
		      \advance\tempc by \wd\tempa
		      \divide\tempc by 2 }

\def\ruleanchelp#1#2#3{\setbox\tempa=%
                            \hbox{$\displaystyle{\mathstrut #2}$}%
			\setbox\tempd=\hbox{$\; #3$}%
		        \setbox\tempb=\vbox{\halign{##\cr
	\mud{#1}\cr
	\noalign{\vskip\the\lineskip}%
	\noalign{\hrule height 0pt}%
	\rig{\vbox to 0pt%
          {\vss\hbox to 0pt{\copy\tempd \hss}\vss}}\cr
	\noalign{\hrule}%
	\noalign{\vskip\the\lineskip}%
	\mud{\copy\tempa}\cr}}%
		      \tempc=\wd\tempb
		      \advance\tempc by \wd\tempa
		      \divide\tempc by 2 }

\def\gaphelp#1#2#3{\setbox\tempa=\hbox{$\displaystyle{\mathstrut #2}$}%
		   \setbox\tempb=\vbox{\lineskip=2pt%
 \halign{##\cr
	\mud{#1}\cr
	\noalign{\hrule height 0pt}%
	\mud{#3}\cr
	\noalign{\hrule height 0pt}%
	\noalign{\vskip\the\lineskip}%
	\mud{\copy\tempa}\cr}}%
		      \tempc=\wd\tempb
		      \advance\tempc by \wd\tempa
		      \divide\tempc by 2 }

\def\inrulean#1#2#3{{\ruleanhelp{#1}{#2}{#3}%
		     \hbox to \wd\tempa{\hss \box\tempb \hss}}}
\def\ginruleanl#1#2#3{{\ruleanchelp{#1}{#2}{#3}%
		       \hbox to .5\linewidth{\hfil
		         \box\tempb\hskip\wd\tempd \hfil}}} 
\def\hinruleanl#1#2#3#4{{\ruleanchelp{#1}{#2}{#3}%
		         \hbox to #4\linewidth{\hfil
		           \box\tempb\hskip\wd\tempd \hfil}}} 

\def\inrulebn#1#2#3#4{\inrulean{#1\quad\qquad #2}{#3}{#4}}

\def\inrulegap#1#2#3{{\gaphelp{#1}{#2}{#3}%
		      \hbox to \wd\tempa{\hss \box\tempb \hss}}}

\title{Polymorphic Lemmas and Definitions in $\lambda$Prolog and Twelf}

\author[Andrew W. Appel and Amy P. Felty]
{ANDREW W. APPEL\\
Department of Computer Science, Princeton University, USA\\
\email{appel@princeton.edu}
\and
AMY P. FELTY\\
School of Information Technology and Engineering, University of
Ottawa, Canada\\
\email{afelty@site.uottawa.ca}}

\begin{document}

\maketitle

\begin{abstract}
\lprolog\ is known to be well-suited for expressing and implementing
logics and inference systems.  We show that lemmas and definitions in
such logics can be implemented with a great economy of expression.  We
encode a higher-order logic using an encoding that maps both terms and
types of the object logic (higher-order logic) to terms of the
metalanguage (\lprolog).  We discuss both the Terzo and Teyjus
implementations of \lprolog.  We also encode the same logic in Twelf
and compare the features of these two metalanguages for our purposes.
\end{abstract}

\section{Introduction}
It has long been the goal of mathematicians to minimize the set of
assumptions and axioms in their systems.  Implementers of theorem
provers use this principle: they use a logic with as few inference
rules as possible, and prove lemmas outside the core logic in
preference to adding new inference rules.  In applications of logic to
computer security -- such as \emph{proof-carrying code}
\cite{necula97} and distributed authentication frameworks
\cite{appel99:says} -- the implementation of the core logic is inside
the trusted code base (TCB), while proofs need not be in the TCB
because they can be checked.

Two aspects of the core logic are in the TCB: a set of
logical connectives and inference rules, and a program in some
underlying programming language that implements proof checking -- that
is, interpreting the inference rules and matching them against
a theorem and its proof.

\begin{figure}
\centerline{\epsfig{file=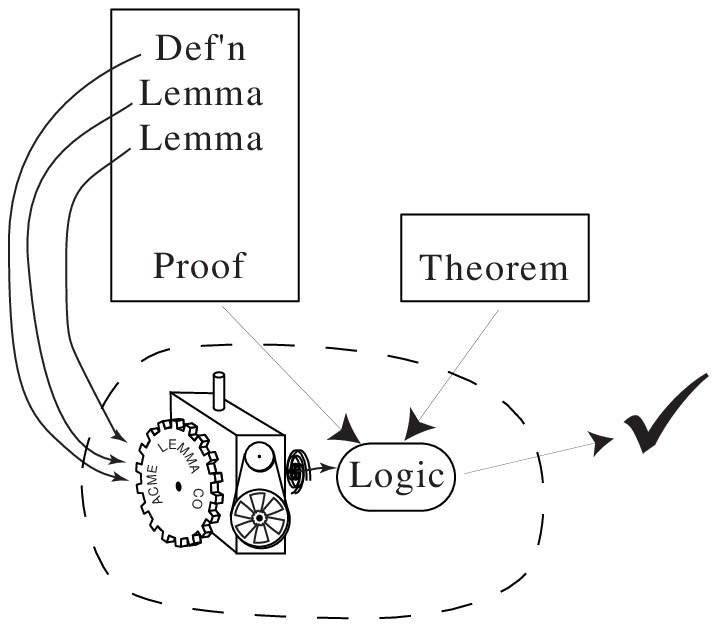}}
\vspace{-18pt}
{\small \hspace{59pt}\hspace{68pt}Trusted code base}
\captionx{Figure}{Lemma machinery is inside the TCB.}
\label{lemmamach}
\end{figure}

Definitions and lemmas are essential in constructing proofs of 
reasonable size and clarity.  A proof system should have machinery
for checking lemmas, and applying lemmas and definitions, in the checking
of proofs.
This machinery also is within the TCB; see Figure~\ref{lemmamach}.
Many theorem provers support definitions and lemmas and provide a
variety of advanced features designed to help with tasks such as
organizing definitions and lemmas into libraries, keeping track of
dependencies, and providing modularization; in our work we are
particularly concerned with separating that part of the machinery
necessary for proof checking (i.e., in the TCB) from the
programming-environment support that is used in proof development.
This separation was particularly important for
a proof-carrying code system we built initially in
\lprolog~\cite{AppelFelty:POPL00}.
In this paper we will demonstrate a definition/lemma implementation
that is about three dozen lines of code.

The \lprolog{} language \cite{Nadathur88a} has several features that
allow concise and clean implementation of logics, proof checkers, and
theorem provers \cite{felty93}.  In a previous
paper~\cite{AppelFelty:ICLP99}, we presented a lemma and definition
mechanism implemented in \lprolog.  In this paper, we extend that work
and describe it more fully.  We present the lemma mechanism and a
generalization of our definition mechanism, again implemented in
\lprolog.  Since we now have more experience using the Twelf
system~\cite{Pfenning91LF,Pfenning99b}, we include a detailed
comparison of the Twelf and \lprolog\ versions of the encoding of our
logic, lemmas, and definitions.
An important purpose of this paper is to show which language features
allow a small TCB and efficient representation of proofs.  We also
give a comparison of programming issues that are important to our
proof-carrying code application.

Although the lemma and definition mechanism is general, we illustrate
it using an implementation of higher-order logic.  We call this logic
the \emph{object logic} to distinguish it from the \emph{metalogic}
implemented by \lprolog{} or Twelf.  Our object logic is not polymorphic, but
our lemma and definition mechanisms are polymorphic in the sense that
they can express properties that hold at any type of the object logic.
The symmetry of equality, for example, is one such lemma we will
encounter.

\section{Encoding a higher-order logic}
\label{core}

The \lprolog{} version of the clauses we present use the syntax of the
Terzo implementation~\cite{Terzo}.  We also discuss the Teyjus
implementation~\cite{NadathurCade99} and compare the two for our
purposes.  Terzo is interpreted and provides more flexibility, but
Teyjus has a compiler in which our code runs much more efficiently.

\lprolog\ is a higher-order logic programming language which extends
Prolog in essentially two ways.  First, it replaces first-order terms
with the more expressive simply-typed $\lambda$-terms; \lprolog\
implementations generally extend simple types to include ML-style
prenex polymorphism \cite{damas82,NadathurPfenning92}.  Second, it
permits implication and universal quantification (over objects of any
type) in goal formulas.

We introduce types and constants using \cd{kind} and \cd{type}
declarations, respectively.  For example, a new primitive type $t$ and
a new constant $f$ of type $t\rightarrow t\rightarrow t$ are declared
as follows.
\begin{verbatim}
kind   t       type.
type   f       t -> t -> t.
\end{verbatim}
Capital letters in type declarations denote type variables and are
used in polymorphic types.  In program goals and clauses,
$\lambda$-abstraction is written using backslash \lamb{} as an infix
operator.  Capitalized tokens not bound by $\lambda$-abstraction
denote free variables.  All other unbound tokens denote constants.
Universal quantification is written using the constant {\tt pi} in
conjunction with a $\lambda$-abstraction (e.g., \cd{pi X\lamb}
represents universal quantification over variable {\tt X}).  The
symbols \emph{comma} and {\tt =>} represent conjunction and
implication.  The symbol {\tt :-} denotes the converse of {\tt =>} and
is used to write the top-level implication in clauses.  The type
\cd{o} is the type of clauses and goals of \lprolog{}.  We usually
omit universal quantifiers at the top level in definite clauses, and
assume implicit quantification over all free variables.

We will encode a natural deduction proof system for our higher-order
object logic.  (In our earlier work~\cite{AppelFelty:ICLP99}, we
implemented a sequent calculus version.)  We implement a proof checker
for this logic that is similar to the one described by
Felty~\cite{felty93}.  Program~\ref{core-types} contains the type
declarations used in our encoding.
\begin{figure}
\begin{verbatim}
kind  tp              type.
kind  tm              type.
kind  pf              type.

type  form            tp.
type  intty           tp.
type  arrow           tp -> tp -> tp.           infixr  arrow   8.
type  pair            tp -> tp -> tp.

type  eq              tp -> tm -> tm -> tm.
type  imp             tm -> tm -> tm.           infixr  imp     7.
type  forall          tp -> (tm -> tm) -> tm.
type  false           tm.

type  lam             (tm -> tm) -> tm.
type  app             tp -> tm -> tm -> tm.
type  mkpair          tm -> tm -> tm.
type  fst             tp -> tm -> tm.
type  snd             tp -> tm -> tm.

type  hastype         tm -> tp -> o.
type  proves          pf -> tm -> o.
type  assump          o -> o.

type  refl            pf.
type  beta            pf.
type  fstpair         pf.
type  sndpair         pf.
type  surjpair        pf.
type  congr           tp -> tm -> tm -> (tm -> tm) -> pf -> pf -> pf.
type  imp_i           (pf -> pf) -> pf.
type  imp_e           tm -> pf -> pf -> pf.
type  forall_i        (tm -> pf) -> pf.
type  forall_e        tp -> (tm -> tm) -> pf -> tm -> pf.
\end{verbatim}
\captionx{Program}{Type declarations for core logic.}
\label{core-types}
\end{figure}

We introduce three primitive types: \cd{tp} for object-level types,
\cd{tm} for object-level terms (including formulas) and \cd{pf} for
proofs in the object logic.

We introduce constants for the object-level type constructors.
The main type constructor for our object language is
the \cd{arrow} constructor taking two types as arguments.
We also include objects of type \cd{tp} to represent base types,
such as \cd{form} and \cd{intty}.

To represent formulas, we introduce constants such as \cd{imp} to
represent implication in the object logic, and \cd{eq} which takes two
terms and a type and is used to represent equality at any type.  We
use infix notation for the type arrow and binary logical connectives.
The binding strength of each infix operator is declared using an
\cd{infix} declaration.  The constant \cd{forall} represents universal
quantification.  It takes a type representing the type of the bound
variable and a functional argument, which allows object-level binding
of variables by quantifiers to be defined in terms of meta-level
$\lambda$-abstraction.  An example of its use is the following
formula, which expresses the commutativity of equality for integers:
\begin{verbatim}
forall intty (X\ forall intty (Y\ (eq intty X Y) imp (eq intty Y X))).
\end{verbatim}
The parser uses the usual rule for the syntactic extent of a lambda,
so this expression is equivalent to
\begin{verbatim}
forall intty X\ forall intty Y\ eq intty X Y imp eq intty Y X.
\end{verbatim}
This use of higher-order data structures is called \emph{higher-order
abstract syntax}~\cite{Pfenning88}; with it, we don't need to describe
the mechanics of substitution explicitly in the object logic
\cite{felty93}.

To represent terms, we introduce the \cd{app} and \cd{lam} constants
for application and abstraction, as well as constants for pairing and
projections.  The \cd{app} constructor takes three arguments.  The
second argument is a term of functional type and the third argument is
the term it is applied to.  The first argument is the type of the
argument to the function.  The \cd{lam} constant has a type, which like
\cd{forall}, uses meta-level abstraction to represent object-level
binding.

The constants at the end of Program~\ref{core-types} are used to build
terms representing proofs.  We call these constants as well as any
other terms whose type ends in ``\cd{-> pf}'' \emph{proof
constructors}.

Programs~\ref{core-types} and~\ref{core-rules} together implement a
full proof checker for our object logic.
\begin{figure}
\begin{verbatim}
hastype (eq T X Y) form :- hastype X T, hastype Y T.
hastype (A imp B) form :- hastype A form, hastype B form.
hastype (forall T A) form :- pi x\ (hastype x T => hastype (A x) form).
hastype false form.
hastype (lam F) (T1 arrow T2) :- pi x\ (hastype x T1 => hastype (F x) T2).
hastype (app T1 F X) T2 :- hastype F (T1 arrow T2), hastype X T1.
hastype (mkpair X Y) (pair T1 T2) :- hastype X T1, hastype Y T2.
hastype (fst T2 X) T1 :- hastype X (pair T1 T2).
hastype (snd T1 X) T2 :- hastype X (pair T1 T2).

proves Q A :- assump (proves Q A).
proves refl (eq T X X).
proves beta (eq T2 (app T1 (lam F) X) (F X)).
proves fstpair (eq T1 (fst T2 (mkpair X Y)) X).
proves sndpair (eq T2 (snd T1 (mkpair X Y)) Y).
proves surjpair (eq (pair T1 T2) (mkpair (fst T2 Z) (snd T1 Z)) Z).
proves (congr T X Z H P1 P2) (H X) :- 
  hastype X T, hastype Z T,
  proves P1 (eq T X Z), proves P2 (H Z).
proves (imp_i Q) (A imp B) :-
  pi p\ (assump (proves p A) => proves (Q p) B).
proves (imp_e A Q1 Q2) B :-
  hastype A form, proves Q1 (A imp B), proves Q2 A.
proves (forall_i Q) (forall T A) :-
  pi y\ (hastype y T => proves (Q y) (A y)).
proves (forall_e T A Q X) (A X) :- 
  pi x\ (hastype x T => hastype (A x) form),
  hastype X T,
  proves Q (forall T A).
\end{verbatim}
\captionx{Program}{Inference rules of the core logic.}
\label{core-rules}
\end{figure}
Program~\ref{core-rules} implements both typechecking and
inference rules.  The last four clauses of Program~\ref{core-rules} implement
the introduction and elimination rules for implication and universal
quantification, which are given in Figure~\ref{core-inf}.
\begin{figure}[tbh]
\vcalign{\ \hspace{-.5in}
         \inrulean{\inrulegap{}{B}{(A)}}{A\supset B}{\supset\mbox{-I}}\cr
	 \hspace{.8in}
         \inrulebn{A}{A\supset B}{B}{\supset\mbox{-E}}\cr
	 \hspace{1.0in}
         \inrulean{\inrulegap{}{[y/x]A}{(y:\tau)}}{\forall_\tau x A}
                  {\forall_\tau\mbox{-I}}\cr
	 \hspace{.8in}
         \inrulebn{\forall_\tau x A}{t:\tau}{[t/x]A}{\forall_\tau\mbox{-E}}\cr}

\centerline{\parbox{4.5in}{The $\forall$-I rule has the proviso
that the variable $y$ cannot appear free in $\forall_\tau x A$, or in any
assumption on which the deduction of $[y/x]A$ depends.}}
\captionx{Figure}{Natural Deduction Inference Rules}
\label{core-inf}
\end{figure}
We do not include inference rules for the other logical connectives.
Instead, we define them in terms of existing connectives using our
definition mechanism described later.
The remaining clauses for the \cd{proves} predicate implement
inference rules for equality.  Typechecking for terms is implemented
by the \cd{hastype} clauses.
Proof checking is implemented by the \cd{proves} clauses.
A goal of the form  \cd{(proves P A)} should be run only
after \cd{A} is typechecked, i.e., a proper check has the form
\cd{(hastype A form, proves P A)}.

To implement the discharge of assumptions in the implication
introduction rule, we use implication and universal quantification in
\lprolog{} goals.  The goal \cd{(D => G)} adds clause \cd{D} to the
\lprolog{} clause database, attempts to solve \cd{G}, and then (upon
either the success or failure of \cd{G}) removes \cd{D} from the
clause database.  The goal \cd{(pi y\lamb (G y))} introduces a new
constant \cd{c} with the same type as \cd{y}, replaces \cd{y} with
\cd{c}, and attempts to solve the goal \cd{(G c)}.  For example,
consider the goal
\begin{verbatim}
proves (imp_i q\q) (a imp a)
\end{verbatim}
where \cd{a} is a propositional constant (a constant of type
\cd{form}); then
\lprolog{} will execute the (instantiated) body of the \cd{imp\_i}
clause
\begin{verbatim}
pi p\ (assump (proves p a) => proves ((q\q) p) a)
\end{verbatim}
This generates a new constant \cd{c}, and adds \cd{(assump (proves c
a)} to the database; then the subgoal \cd{(proves
((q\lamb{}q) c) a)}, which is $\beta$-equivalent to
\cd{(proves c a)}, matches the first clause for the
\cd{proves} predicate.  The subgoal \cd{(assump (proves c
a))} is generated and this goal matches our dynamically added clause.
We have chosen to use the \cd{assump} predicate for adding atomic
clauses to the program.  This is not necessary, but we find it useful
to distinguish between adding atomic clauses and adding non-atomic
clauses, which we will see later.
Note that the typechecking clauses for \cd{forall} and \cd{lam} use
meta-level implication and universal quantification in a manner
similar to the \cd{proves} clause for the $\supset$-I rule.

It is important to show that our encoding of higher-order logic in
\lprolog\ is {\em adequate}.  To do so, we must show that a formula
has a natural deduction proof if and only if its representation as a
term has an associated proof term that can be checked using the
inference rules of Program~\ref{core-rules}.  The encoding we use is
similar to the encoding of higher-order logic in the Logical
Framework~\cite{Harper89} and the proof of adequacy of our encoding is
similar to the one discussed there.  The main difference between the
two encodings is the types of the logical connectives.  For example,
in their encoding, \cd{imp} is given type \cd{tm} and the fact that it
is a connective which takes two formulas as arguments is expressed
using object level types; the \cd{hastype} clause is
\begin{verbatim}
hastype imp (form arrow form arrow form).
\end{verbatim}
An implication must then be expressed using the \cd{app} constructor,
e.g., \cd{(app (app imp A) B)}.  We found that this encoding of the
connectives quickly became cumbersome and our encoding was more
readable.  On the other hand, our encoding is not as economical as the
one we used previously~\cite{AppelFelty:ICLP99}.  There we represented
object-level types as meta-level types, which allowed us to eliminate
all the \cd{hastype} clauses and subgoals.  The types of our object
logic, however, did not match up well with the types of \lprolog,
which forced certain limitations in the implementation of our
proof-carrying code system.  (See Appel and
Felty~\cite{AppelFelty:ICLP99} for further
analysis.)  The encoding in the current paper seems to be the best
compromise.

\section{Lemmas}
\label{lemmas}
In mathematics the use of lemmas can make a proof more readable by
structuring the proof, especially when the lemma corresponds to some
intuitive property.  For automated proof checking (in contrast to
automated or traditional theorem proving) this use of lemmas is not
essential, because the computer doesn't need to understand the proof
in order to check it.  But lemmas can also reduce the \emph{size} of a
proof (and therefore the time required for proof checking): when a
lemma is used multiple times it acts as a kind of ``subroutine.''
This is particularly important in applications like proof-carrying
code where proofs are transmitted over networks to clients who check
them.
We first present an example which we use to illustrate our lemma
mechanism in \lprolog\ (Section~\ref{lemmasExample}), and then present
this mechanism as we'd implement it in Terzo
(Section~\ref{lemmasTerzo}).  We then explain the modifications
required to meet the extra restrictions imposed by Teyjus
(Section~\ref{lemmasTeyjus}).  We end this section with some
optimizations that are important for keeping proofs that use lemmas as
small as possible (Section~\ref{lemmasOpt}) and then with some more
examples (Section~\ref{lemmasMore}).

\subsection{An example}
\label{lemmasExample}

Theorem~\ref{thm:simple} shows the use of our core logic to express a
simple proof checking goal.
\begin{figure}
\begin{verbatim}
proves
  (forall_i I\ (forall_i J\ (imp_i Q\
    (congr intty I J (eq intty J) Q refl))))
  (forall intty I\ forall intty J\ (eq intty I J imp eq intty J I)).
\end{verbatim}
\captionx{Theorem}{$\forall_\mathit{int}\; I\; \forall_\mathit{int}\; J\; 
((I=_\mathit{int}\;J) \supset (J=_\mathit{int}\;I))$.}
\label{thm:simple}
\end{figure}
The proof of this lemma uses the $\forall$-I rule as well as
congruence and reflexivity of equality.  Its proof can be checked as a
successful \lprolog{} query to our core logic in
Programs~\ref{core-types} and \ref{core-rules}.  Alternatively, we may
want to prove it using the following general lemma about symmetry of
equality at any type.
$$\inrulean{A:\tau\quad B:\tau\quad B =_\tau A}{A =_\tau B}{}$$
The proof of this lemma can be checked as the following \lprolog{}
query.
\begin{verbatim}
pi T\ pi A\ pi B\ pi P\
  (hastype A T, hastype B T, proves P (eq T B A)) => 
    proves (congr T B A (eq T A) P refl) (eq T A B).
\end{verbatim}
This query introduces an arbitrary \cd{P}, adds the typing clauses
\cd{(hastype A T)} and \cd{(hastype B T)}, and the assumption
\cd{(proves P (eq T B A))} to the set of clauses,
then checks the proof of congruence using these facts.
The syntax $F$ \cd{=>} $G$ means exactly the same as $G$ \cd{:-} $F$, so
we could just as well write this query as
\begin{verbatim}
pi T\ pi A\ pi B\ pi P\
  (proves (congr T B A (eq T A) P refl) (eq T A B) :-
     hastype A T, hastype B T, proves P (eq T B A)).
\end{verbatim}

Now, suppose we abstract the proof (roughly, \cd{congr T B A (eq T A)
P refl}) from this query.
\begin{verbatim}
(Inference = (PCon\ pi T\ pi A\ pi B\ pi P\ 
               proves (PCon T A B P) (eq T A B) :-
                 hastype A T, hastype B T, proves P (eq T B A)),
 Proof = (T\A\B\P\ congr T B A (eq T A) P refl),
 Query = (Inference Proof),
 Query).
\end{verbatim}
The solution of this query proceeds in four steps: the variable 
\cd{Inference} is unified with a $\lambda$-term; \cd{Proof} is unified
with a $\lambda$-term; \cd{Query} is unified with the application of
\cd{Inference} to \cd{Proof} (which is a term $\beta$-equivalent to
the query of the previous paragraph), and finally \cd{Query} is solved
as a goal (checking the proof of the lemma).

Once we know that the lemma is valid, we can make a new \lprolog{}
atom \cd{symm} to stand for its proof, and we can prove some other
theorem in a context where the clause \cd{(Inference symm)} is in the
clause database; remember that \cd{(Inference symm)} is
$\beta$-equivalent to
\begin{verbatim}
pi T\ pi A\ pi B\ pi P\
 (proves (symm T A B P) (eq T A B) :-
   hastype A T, hastype B T, proves P (eq T B A)).
\end{verbatim}
This series of transformations starting with a proof checking subgoal
has led us to a clause that looks remarkably like an inference rule.
With this clause in the database, we can use the new proof constructor
\cd{symm} just as if it were primitive.  Instead of adding new clauses
like this to our proof checker, which would increase the size of our
TCB, we show how to put such lemmas inside proofs.

\subsection{Lemmas in proofs}
\label{lemmasTerzo}

In the example in the previous section, \cd{symm} is a new constant,
but when lemmas are proved and put inside proofs dynamically, we can
instead ``make a new atom'' by simply \cd{pi}-binding it.  This leads
to the recipe for lemmas shown in Program~\ref{lemmafig}, which is the
heart of our lemma mechanism.
\begin{figure}
\begin{verbatim}
type  lemma_pf  (A -> o) -> A -> (A -> pf) -> pf.

proves (lemma_pf Inference LemmaProof RestProof) C :-
 Inference LemmaProof,
 pi Name\ ((Inference Name) => (proves (RestProof Name) C)).
\end{verbatim}
\captionx{Program}{The \texttt{lemma\_pf} proof constructor.}
\label{lemmafig}
\end{figure}
(We will improve it slightly in the next section.)  This program
introduces a constructor \cd{lemma\_pf} for storing lemmas in proofs.
This constructor takes three arguments: (1) a derived inference rule
\cd{Inference} (of type \cd{A -> o}) parameterized by a proof
constructor (of type \cd{A}), (2) a term \cd{LemmaProof} of type
\cd{A} representing a proof of the lemma built from core-logic proof
constructors (or using other lemmas), and (3) a proof of the main
theorem \cd{RestProof} that is parameterized by a proof constructor
(of type \cd{A}).  Operationally, this clause first executes
\cd{(Inference LemmaProof)} as a query, to check the proof of the
lemma itself; then it \cd{pi}-binds \cd{Name} in the lemma, adds it as
a new clause, and runs \cd{RestProof} (which is parameterized on the
lemma proof constructor) applied to \cd{Name}.

The terms \cd{Inference} and \cd{Proof} from the example in
Section~\ref{lemmasExample} illustrate the form of the terms which
will appear as the first two arguments to \cd{lemma\_pf}.
Theorem~\ref{thm:symm} illustrates the use of \cd{lemma\_pf} in an
example; this theorem is a modification of Theorem~\ref{thm:simple}
that uses the \cd{symm} lemma.
\begin{figure}
\begin{verbatim}
proves
(lemma_pf
  (Symm\ pi T\ pi A\ pi B\ pi P\
    proves (Symm T A B P) (eq T A B) :-
        hastype A T, hastype B T, proves P (eq T B A))
  (T\A\B\P\ (congr T B A (eq T A) P refl))
  (symm\ (forall_i I\ (forall_i J\ (imp_i Q\ (symm intty J I Q))))))
(forall intty I\ forall intty J\ (eq intty I J imp eq intty J I)).
\end{verbatim}
\captionx{Theorem}{Modification of Theorem~\ref{thm:simple} to use a lemma.}
\label{thm:symm}
\end{figure}

\subsection{Lemmas in Teyjus}
\label{lemmasTeyjus}

If we restrict ourselves to the Terzo implementation of \lprolog{},
then meta-level formulas can occur inside proofs using any of the
\lprolog{} connectives.  But if we want to be able to use
Teyjus as well, we must make one more change.
The Teyjus system does not allow \cd{=>} or \cd{:-} to appear in
arguments of predicates.  Thus the term
\begin{verbatim}
  (Symm\ pi T\ pi A\ pi B\ pi P\
    proves (Symm T A B P) (eq T A B) :-
        hastype A T, hastype B T, proves P (eq T B A))
\end{verbatim}
occurring in the \cd{symm} lemma in Theorem~\ref{thm:symm} cannot
appear directly as the first argument to \cd{lemma\_pf}.  Teyjus also
does not allow variables to appear at the head of the left of an
implication.  These restrictions come from the theory underlying
\lprolog~\cite{Miller91apal}; without the latter one, a runtime check
is needed to insure that every dynamically created goal is an
acceptable one.

We can avoid putting \cd{:-} inside arguments of predicates by writing
the above term as
\begin{verbatim}
  (Symm\ pi T\ pi A\ pi B\ pi P\
    proves (Symm T A B P) (eq T A B) <<==
        hastype A T, hastype B T, proves P (eq T B A))
\end{verbatim}
where \cd{<<==} is a new infix operator of type \cd{o -> o}.  But
this, in turn, means that the subgoal \cd{(Inference LemmaProof)} of
the \cd{lemma\_pf} clause in Program~\ref{lemmafig} will no longer
check the lemma, since \cd{<<==} has no operational meaning.  To
handle such goals, we add the three constants declared at the
beginning of Program~\ref{cinterp}, which introduce both
forward and backward implication arrows, and a new atomic predicate
\cd{cl} of type \cd{o -> o}, and we introduce the two clauses that
follow these declarations to interpret our new arrows as \lprolog{}
implication.
\begin{figure}
\begin{verbatim}
type  ==>>            o -> o -> o.              infixr  ==>>    4.
type  <<==            o -> o -> o.              infixl  <<==    0.
type  cl              o -> o.

(D ==>> G) :- (cl D) => G.
(G <<== D) :- (cl D) => G.

type  backchain    o -> o -> o.

proves P A :- cl Cl, backchain (proves P A) Cl.
hastype X T :- cl Cl, backchain (hastype X T) Cl.
assump G :- cl Cl, backchain (assump G) Cl.

backchain G G.
backchain G (pi D) :- backchain G (D X).
backchain G (A,B) :-  backchain G A; backchain G B.
backchain G (H <<== G1) :-  backchain G H, G1.
backchain G (G1 ==>> H) :-  backchain G H, G1.
\end{verbatim}
\captionx{Program}{An interpreter for dynamic clauses.}
\label{cinterp}
\end{figure}
Note that although it would have been more direct, we did not add:
\begin{verbatim}
(D ==>> G) :- D => G.
\end{verbatim}
because of the Teyjus restriction mentioned above that variables
cannot appear at the head of the left of an implication.  The use of
the \cd{cl} ``wrapper'' solves the problem created by this
restriction, but requires us to implement an interpreter to handle
clauses of the form \cd{(cl A)}.  The remaining clauses in
Program~\ref{cinterp} implement this interpreter.

Since the type of \cd{(Inference Proof)} is \cd{o}, the term
\cd{Inference} might conceivably contain subterms which are \lprolog\
clauses.  Of course, in Teyjus these clauses will not contain \cd{:-} or
\cd{=>}, but they may contain \cd{<<==} and \cd{==>>}, which get
interpreted via the clauses of Program~\ref{cinterp}.  They could
also, for example, contain any other \lprolog\ code including
input/output operations.  Executing \cd{(Inference Proof)} cannot lead
to unsoundness -- if the resulting proof checks, it is still valid.
But there are some contexts where we wish to restrict the kind of
program that can occur inside a proof and be run when the proof is
checked.  For example, in a proof-carrying-code system, the code
consumer might not want proof checking to cause \lprolog{} to execute
code that accesses private local resources.

To limit the kind and amount of execution possible in the 
executable part of a lemma, we introduce the \cd{valid\_clause}
predicate of type \cd{o -> o} (Program~\ref{validclause}).
\begin{figure}
\begin{verbatim}
valid_clause (pi C) :-     pi X\ valid_clause (C X).
valid_clause (A,B) :-      valid_clause A, valid_clause B.
valid_clause (A <<== B) :- valid_clause A, valid_clause B.
valid_clause (A ==>> B) :- valid_clause A, valid_clause B.
valid_clause (proves Q A).
valid_clause (hastype X T).
valid_clause (assump (proves Q A)).
\end{verbatim}
\captionx{Program}{Valid clauses.}
\label{validclause}
\end{figure}
A clause is valid if it contains \cd{pi}, \emph{comma}, \cd{<<==}, \cd{==>>},
\cd{proves}, \cd{hastype}, \cd{assump}, and nothing else.  Of course,
a \cd{proves} or \cd{assump} clause contains subexpressions of type
\cd{pf} and \cd{tm}, and a \cd{hastype} clause has subexpressions of
type \cd{tm} and \cd{tp}, so all the constants in proofs, terms, and
types of our object logic are also permitted.  Absent from this list
are \lprolog{} input/output (such as \cd{print}) and the semicolon
(backtracking search).

The \cd{valid\_clause} restriction is the reason that we only need new
clauses for the \cd{proves}, \cd{hastype}, and \cd{assump} predicates
in Program~\ref{cinterp}.  We must add at least these three because
they are used for checking nodes in a proof that require using the
clauses added dynamically via the \cd{cl} predicate.  Including no
other predicates in the \cd{valid\_clause} definition guarantees that
we need no other new clauses with \cd{cl} subgoals.

Because of the introduction of \cd{<<==}, \cd{==>>}, and
\cd{valid\_clause}, we modify the clause in
Program~\ref{lemmafig} for checking lemmas.  The new clause is shown
in Program~\ref{teyjuslemmafig}.
\begin{figure}
\begin{verbatim}
proves (lemma_pf Inference LemmaProof RestProof) C :-
 pi Name\ (valid_clause (Inference Name)),
 Inference LemmaProof,
 pi Name\ (cl (Inference Name) => (proves (RestProof Name) C)).
\end{verbatim}
\captionx{Program}{The clause for lemmas in Teyjus.}
\label{teyjuslemmafig}
\end{figure}
The first subgoal is new; it \cd{pi}-binds \cd{Name} and checks to see
if the new lemma applied to \cd{Name} is valid.  The only other
modification is in the last subgoal, which adds the lemma as a new
clause via the \cd{cl} predicate.  Since all lemmas will be added via
\cd{cl}, the only way to use them is via the \cd{proves} clause in
Program~\ref{cinterp}.  Using that clause, the \cd{(cl Cl)} subgoal
looks up the lemmas that have been added, one at a time, and tries
them out via the \cd{backchain} predicate.  This predicate processes
the clauses in a manner similar to the \lprolog{} language itself.  In
Terzo, using this interpreter is less efficient than the direct
implementation in Program~\ref{lemmafig}.  In Teyjus, the interpreter
is required, but when compiled, the code runs faster than either Terzo
version.

In summary, our technique allows lemmas to be contained \emph{within}
the proof.  We do not need to install new ``global'' lemmas into the
proof checker.  The dynamic scoping also means that the lemmas of one
proof cannot interfere with the lemmas of another, even if they have
the same names.  This machinery uses several interesting features of
\lprolog:
\paragraph{Polymorphism.}
The type of the \cd{lemma\_pf} constructor uses polymorphism to indicate
that proof constructors introduced for lemmas can have different
types.
\paragraph{Meta-level formulas as terms.}
Lemmas such as symmetry of equality occur inside proofs as an argument
to the \cd{lemma\_pf} constructor in the following form.
\begin{verbatim}
  (Symm\ pi T\ pi A\ pi B\ pi P\
    proves (Symm T A B P) (eq T A B) <<==
        hastype A T, hastype B T, proves P (eq T B A))
\end{verbatim}
It is just a data structure (parameterized by \cd{Symm}); it does not
``execute'' anything, in spite of the fact that it contains the
\lprolog{} quantifier \cd{pi} and our new connective \cd{<<==}.  This
gives us the freedom to write lemmas using syntax very similar to that
used for writing primitive inference rules.  Handling the new
constants for \cd{<<==} and \cd{==>>} is easy enough operationally.
However, it is an inconvenience for the user, who must use different
syntax in lemmas than in inference rules.  This inconvenience is
avoided in Terzo.
\paragraph{Dynamically constructed goals.}
When the clause from Program~\ref{teyjuslemmafig} for the
\cd{lemma\_pf} proof constructor checks the proof of a lemma by
executing the goal \cd{(Inference LemmaProof)}, we are executing a
goal that is built from a run-time-constructed data structure.
\cd{Inference} will be instantiated with terms such as the one above
representing the symmetry lemma.  It is only when such a term is
applied to its proof and thus appears in ``goal position'' that it
becomes the current subgoal on the execution stack.
\paragraph{Dynamically constructed clauses.}
When, having successfully checked
the proof of a lemma, the \cd{lemma\_pf} clause executes 
\begin{verbatim}
cl (Inference Name) => (proves (RestProof Name) C)
\end{verbatim}
it is adding a dynamically constructed clause to the \lprolog{}
database.

Although it is not the case for Terzo or Teyjus, if a metalanguage
were to prohibit all terms having \cd{o} in their types as arguments
to a predicate, it would still be possible to implement lemmas using
our approach.  Appendix~\ref{dynamic} illustrates by showing an
interpreter which extends Program~\ref{cinterp} to handle this extra
restriction.  New constants must be introduced not only for
implication but also for every meta-level connective.  Note that when
meta-level formulas are not allowed, there is no possibility for
dynamically created goals or clauses.  Twelf for example, does not
allow meta-level formulas as terms and is also not polymorphic, and
thus the approach described in this section cannot be used, but the
approach of Appendix~\ref{dynamic} could.  Instead, as we will see in
Section~\ref{sec:twelf}, Twelf provides alternative features which we
can use to implement lemmas.

\subsection{Some optimizations for implementing lemmas}
\label{lemmasOpt}

The \cd{Symm} proof constructor in Theorem~\ref{thm:symm} is a bit
unwieldy, since it requires \cd{T}, \cd{A}, and \cd{B} as arguments.
We can imagine writing a primitive inference rule
\begin{verbatim}
proves (symm P) (eq T A B) :-
  hastype A T, hastype B T, P proves (eq T B A).
\end{verbatim}
using the principle that the proof checker doesn't need to be told
\cd{T}, \cd{A}, and \cd{B} inside the proof term, since they can be
found in the formula to be checked.  Then, in Theorem~\ref{thm:symm},
\cd{(Symm intty J I Q)} would be \cd{(Symm Q)}.

Therefore we add three new proof constructors---\cd{elam}, \cd{extract},
and \cd{extract\-Goal}\----as shown in Program~\ref{elam}.
\begin{figure}
\begin{verbatim}
type  elam            (A -> pf) -> pf.
type  extract         tm -> pf -> pf.
type  extractGoal     o -> pf -> pf.

proves (elam Q) A :- proves (Q B) A.
proves (extract A P) A :- proves P A.
proves (extractGoal G P) A :- valid_clause G, G, proves P A.
\end{verbatim}
\captionx{Program}{Proof constructors for implicit arguments of lemmas.}
\label{elam}
\end{figure}
These can be used in the following stereotyped way to extract components
of the formula to be proved.  First bind variables with \cd{elam}, then
match the target formula with \cd{extract}.  Theorem~\ref{thm:symma} is a
modification of Theorem~\ref{thm:symm} that makes use of these constructors.
\begin{figure}
\begin{verbatim}
proves
(lemma_pf
  (Symm\ pi T\ pi A\ pi B\ pi P\
    proves (Symm P) (eq T A B) <<== 
        hastype A T, hastype B T, proves P (eq T B A))
  (P\ elam T\ elam A\ elam B\
    (extract (eq T A B) (congr T B A (eq T A) P refl)))
  (symm\ (forall_i I\ (forall_i J\ (imp_i Q\ (symm Q))))))
(forall intty I\ forall intty J\ (eq intty I J imp eq intty J I)).
\end{verbatim}
\captionx{Theorem}{$\forall_\mathit{int}\; I\; \forall_\mathit{int}\; J\; 
((I=_\mathit{int}\;J) \supset (J=_\mathit{int}\;I))$.}
\label{thm:symma}
\end{figure}

Note that we could eliminate the \cd{hastype} subgoals from our new
version of the \cd{symm} lemma because we know them to be redundant as
long as \cd{(eq T A B)} was already typechecked.  The reason for
keeping them is that the second subgoal of the clause in
Program~\ref{teyjuslemmafig} would fail without them; the proof checking of
the lemma requires these \cd{hastype} assumptions.  In encoding our
core logic, it was possible to eliminate all such redundant subgoals.
The fact that such a shortcut is not possible in lemmas causes a
tradeoff; by keeping such lemmas out of the TCB and
putting them in proofs, we are forcing the proof checker to do more
work.  There seems to be no easy way to avoid this redundant work,
though some ad-hoc optimizations to proof checking might be possible.

The \cd{extractGoal} proof constructor asks the checker to run
\lprolog{} code to help construct the proof.  Its implementation uses
\cd{valid\_clause} to restrict what kinds of \lprolog{} code can be
run.  Note, however, that \cd{valid\_clause} does not always eliminate
code that loops and so its current implementation cannot guarantee
termination.  A stricter \cd{valid\_clause} would be necessary to
achieve this.

The \cd{extractGoal} proof constructor was useful for handling
assumptions in the sequent calculus version of our object
logic~\cite{AppelFelty:ICLP99}; for natural deduction, the same need
does not arise in the implementation of our core logic, but
\cd{extractGoal} is useful for implementing more complex lemmas.
Although we have not done so, it would be interesting to further
explore the possibility of creating more compact proofs by leaving out
information that can be computed easily via code given as arguments to
\texttt{extractGoal}.

\subsection{More examples}
\label{lemmasMore}

As another example of the use of lemmas, we can of course use one
lemma in the proof of another, as shown by
Theorem~\ref{thm:symmtrans}.
\begin{figure}
\begin{verbatim}
proves
(lemma_pf
  (Symm\ pi T\ pi A\ pi B\ pi P\
    proves (Symm P) (eq T A B) <<== 
        hastype A T, hastype B T, proves P (eq T B A))
  (P\ elam T\ elam A\ elam B\
    (extract (eq T A B) (congr T B A (eq T A) P refl)))
  (symm\
(lemma_pf
  (Trans\ pi T\ pi A\ pi B\ pi C\ pi Q1\ pi Q2\
    proves (Trans C Q1 Q2) (eq T A B) <<==
        hastype A T, hastype B T, hastype C T,
        proves Q1 (eq T A C), proves Q2 (eq T C B))
  (C\Q1\Q2\ elam A\ elam B\ elam T\
    (extract (eq T A B) (congr T B C (eq T A) (symm Q2) Q1)))
  (trans\ (forall_i I\ forall_i J\ forall_i K\
            (imp_i Q1\ (imp_i Q2\ (trans J (symm Q1) Q2))))))))
(forall intty I\ forall intty J\ forall intty K\
  (eq intty J I imp eq intty J K imp eq intty I K))).
\end{verbatim}
\captionx{Theorem}{$\forall_\mathit{int}\;\; I,J,K\;
((J=_\mathit{int}\;I)\supset(J=_\mathit{int}\;K)
\supset (I=_\mathit{int}\;K))$.}
\label{thm:symmtrans}
\end{figure}
The proof of the \cd{trans} lemma expressing transitivity of equality
uses the \cd{symm} lemma.

The \cd{symm} lemma is naturally polymorphic: it can express the
idea that $(a=_\mathit{int}\;3)\supset(3=_\mathit{int}\;a)$ just as
well as $(f=_\mathit{int\rightarrow int}\;\lambda x.3)\supset(\lambda
x.3=_\mathit{int\rightarrow int}\;f)$.  Theorem~\ref{thm:compose}
illustrates part of a proof which contains two lemmas whose proofs
use \cd{symm} at different types.
\begin{figure}
\begin{verbatim}
(lemma_pf
  (Symm\ pi T\ pi A\ pi B\ pi P\
    proves (Symm P) (eq T A B) <<== 
        hastype A T, hastype B T, proves P (eq T B A))
  (P\ elam T\ elam A\ elam B\
    (extract (eq T A B) (congr T B A (eq T A) P refl)))
  (symm\
(lemma_pf
  (Poly1\ proves Poly1
    (forall (intty arrow intty) f\ forall (intty arrow intty) g\
     (eq (intty arrow intty) f g) imp (eq (intty arrow intty) g f)))
  (forall_i f\ (forall_i g\ (imp_i q\ (symm q))))
  (poly1\
(lemma_pf
  (Poly2\ proves Poly2
     (forall (intty arrow intty) f\ forall intty x\
      (eq intty (app intty f x) x) imp (eq intty x (app intty f x))))
  (forall_i f\ (forall_i x\ (imp_i q\ (symm q))))
  (poly2\ ...))))))
\end{verbatim}
\captionx{Theorem}{Proof with lemmas:
$\forall_\mathit{int\rightarrow int}\;\; f,g\;
((f=_\mathit{int\rightarrow int}\;g) \supset
 (g=_\mathit{int\rightarrow int}\;f))$ and
$\forall_\mathit{int\rightarrow int}\; f\; \forall_\mathit{int}\;x\;
((f(x)=_\mathit{int}\;x) \supset (x=_\mathit{int}\;f(x)))$.}
\label{thm:compose}
\end{figure}
In our previous work~\cite{AppelFelty:ICLP99}, because we represented
object-level types as meta-level types, we were unable to allow
polymorphism in lemmas at all.  To do so would have required a
metalanguage with more general non-prenex polymorphism.  To handle
Theorem~\ref{thm:compose} required two copies of the \cd{symm} lemma,
one at each type.

In principle, we do not need lemmas at all.  Instead, we can replace
each subproof of the form \cd{(lemma\_pf I L R)} with the term \cd{(R L)},
which replaces each use of a lemma with its proof.  This approach,
however, adds undesirable complexity to proofs.  But, using this fact
it should be
straightforward to prove the correspondence between proofs with the
\cd{lemma\_pf} constructor and proofs without, which would directly extend
soundness and adequacy results to our system with lemmas.

\section{Definitions}
\label{defs}

Definitions are another important mechanism for structuring proofs to
increase clarity and reduce size.  If some property (of a base-type
object, or of a higher-order object such as a predicate) can be
expressed as a logical formula, then we allow the introduction of an
abbreviation to stand for that formula.

We start by presenting a motivating example
(Section~\ref{defsExample}), which leads us to our definition
mechanism in \lprolog\ (Section~\ref{defsTeyjus}).  We also discuss
two simpler versions of our definition mechanism
(Sections~\ref{defsSimple} and~\ref{defsAtomic}), which allow us to
have a smaller TCB, but which require more work to use.

\subsection{A motivating example}
\label{defsExample}

We can express the fact that $f$ is an associative function by the
formula
$$\forall_\tau\; X, Y, Z\; (f\,X\,(f\,Y\,Z)=_\tau f\,(f\,X\,Y)\,Z).$$
This will only be a valid expression if $f$ has type
$\tau\rightarrow\tau\rightarrow\tau$.  Putting this formula in
\lprolog{} notation and expressing the type constraint on $f$, we get
the following provable \lprolog{} typechecking goal.
\begin{verbatim}
pi F\ pi T\
 (pi X\ pi Y\ hastype X T => hastype Y T => hastype (F X Y) T) =>
 hastype (forall T X\ forall T Y\ forall T Z\
          eq T (F X (F Y Z)) (F (F X Y) Z)) form.
\end{verbatim}
To make this into a definition, the first step is
to associate some name, say
\cd{assoc}, with the definition body (which is the first argument of
the last \cd{hastype} above).
We associate a name to a body of a
definition in the same way we associated a new proof constructor with
the proof it stood for.  If we follow exactly the pattern of the
\cd{symm} lemma introduced at the beginning of Section~\ref{lemmas},
we abstract out the body of the definition and obtain the following
query.
\begin{verbatim}
(TypeInf = (assoc\ pi F\ pi T\
            hastype (assoc F T) form <<==
            pi X\ pi Y\ (hastype X T ==>>
               hastype Y T ==>> hastype (F X Y) T)),
Def = (F\T\ (forall T X\ forall T Y\ forall T Z\
        (eq T (F X (F Y Z)) (F (F X Y) Z)))),
Query = (TypeInf Def),
Query)
\end{verbatim}
\cd{TypeInf} is the typechecking query above with \cd{=>} replaced by
\cd{==>>} or \cd{<<==}, and the abstraction
\cd{assoc} replacing the body of the definition.
\cd{Def} contains the body abstracted
with respect to the function \cd{F} and type \cd{T} and \cd{(TypeInf
Def)} is exactly the typechecking subgoal above
(except for the use of \cd{==>>} and \cd{<<==}).
If all we wanted
was a typechecking lemma to typecheck expressions of the form given by
\cd{Def}, then we could use our lemma mechanism directly.
\begin{verbatim}
(lemma_pf
  (Assoc\ pi F\ pi T\
    hastype (Assoc F T) form <<==
    pi X\ pi Y\ (hastype X T ==>> hastype Y T ==>> hastype (F X Y) T))
  (F\T\ (forall T X\ forall T Y\ forall T Z\
   (eq T (F X (F Y Z)) (F (F X Y) Z))))
  (assoc\ ...
\end{verbatim}
This example shows that we can have typechecking lemmas in addition to
proof checking lemmas.  It also motivates our definition mechanism
shown next, which we obtain by adding the ability to replace a name
with the expression it represents and vice versa.

\subsection{Implementing definitions}
\label{defsTeyjus}

We introduce a new proof constructor \cd{def\_pf} and a new proof term
\cd{def} to represent equality between a name and its definition.
This definition mechanism is implemented by the clauses in
Program~\ref{definition-fig}.
\begin{figure}
\begin{verbatim}
type  def_pf          tp -> (A -> o) -> A -> (A -> pf) -> pf.
type  def             pf.
type  def_to_eqclause tp -> A -> A -> o -> o.

def_to_eqclause T DName Def (pi Clause) :-
  pi x\ (def_to_eqclause T (DName x) (Def x) (Clause x)).
def_to_eqclause T DName Def (proves def (eq T DName Def)).

proves (def_pf T TypeInf Term RestProof) C :-
  pi Name\
   (valid_clause (TypeInf Name),
    TypeInf Term,
    def_to_eqclause T Name Term (EqClause Name),
    cl (TypeInf Name) => cl (EqClause Name) => (proves (RestProof Name) C)).
\end{verbatim}
\captionx{Program}{Machinery for definitions.}
\label{definition-fig}
\end{figure}
The arguments to \cd{def\_pf} are similar to the arguments to
\cd{lemma\_pf}, but also include one more for the type of the body of
the definition (after it is applied to all its arguments).  In the
clause for proof checking \cd{def\_pf} nodes, the first two subgoals
are similar to \cd{lemma\_pf} nodes.  Here, they check that the
typechecking clause is valid and that \cd{Term} (the body of the
definition) is correctly typed.  The third clause computes the clause
for expressing definitional equality using the \cd{def\_to\_eqclause}
program.  The fourth subgoal for proof checking definitions adds both
the typechecking clause and the equality clause before checking the
rest of the proof.

Like ML, \lprolog{} has parametric polymorphism (in the syntactic
sense).  But unlike ML, \lprolog{} does not have the parametricity
property.  A polymorphic function can examine the structure of its
argument.  We illustrate with a simple example: a function that tells
the arity (number of function arguments) of an arbitrary value.
\begin{verbatim}
type    arity           A -> int -> o.
arity F N :- arity (F X) N1, N is N1 + 1.
arity X 0.
\end{verbatim}
The first clause can only be used when \cd{F} is a function; the second
clause matches any value.
The \cd{def\_to\_eqclause} clauses uses this exact feature of \lprolog's
polymorphism.  It first uses the meta-level type of \cd{Def} to apply
\cd{Def} to as many arguments as possible.  The first clause
introduces new variables to serve as these arguments.  Once it is
applied to all of its arguments, the second clause forms the equality
clause using the type, the name, and the body of the definition.  For
our example, the computed clause is
\begin{verbatim}
EqClause = (assoc\ (pi F\ pi T\
             proves def (eq form (assoc F T)
               (forall T X\ forall T Y\ forall T Z\
                 (eq T (F X (F Y Z)) (F (F X Y) Z)))))).
\end{verbatim}
To ensure that there is only one solution to the \cd{arity} predicate
above and likewise the \cd{def\_to\_eqclause} predicate in
Program~\ref{definition-fig}, we could have used the logic programming
cut (\cd{!}) operator at the end of the first clause for each
predicate.  We have omitted it here because \cd{def\_to\_eqclause} is
only be used in our proof checker, which is written to avoid the need
for backtracking.

To use definitions in proofs we introduce two new lemmas:
\texttt{def\_i} to replace a formula with the definition that stands
for it (or viewed in terms of backward proof, to replace a defined
name with the term it stands for), and \texttt{def\_e} to expand a
definition in the forward direction during proof construction.  Their
proofs are shown in Program~\ref{defs-fig}.
\begin{figure}
\begin{verbatim}
(lemma_pf
  (Def_i\ pi T\ pi Name\ pi B\ pi P\ pi Q1\ pi Q2\
    proves (Def_i T Name B P Q1 Q2) (P Name) <<==
     proves Q1 (eq T Name B),
     hastype Name T, hastype B T,
     proves Q2 (P B))
  (T\Name\B\P\Q1\Q2\ (congr T Name B P Q1 Q2))
  (def_i\
(lemma_pf
  (Def_e\ pi T\ pi Name\ pi B\ pi P\ pi Q1\ pi Q2\
    proves (Def_e T Name B P Q1 Q2) (P B) <<==
     proves Q1 (eq T Name B),
     hastype Name T, hastype B T,
     proves Q2 (P Name))
  (T\Name\B\P\Q1\Q2\ (congr T B Name P
   (congr T Name B (eq T B) Q1 refl) Q2))
  (def_e\...
\end{verbatim}
\captionx{Program}{Lemmas for folding and unfolding definitions.}
\label{defs-fig}
\end{figure}
Theorem~\ref{thm:def} shows a proof using definitions.  In this proof,
\cd{f} is a function symbol and \cd{t} is a type, and the theorem is
represented as a \lprolog{} subgoal with a top-level implication,
where the right hand side is a \cd{proves} subgoal and the left hand
side specifies the typing information about \cd{f} which must hold in
order for the proof in the \cd{proves} subgoal to be valid.
\begin{figure}
\begin{verbatim}
pi f\ pi t\
  (pi x\ pi y\ hastype x t => hastype y t => hastype (f x y) t) =>
(proves
 (lemma_pf ... symm\
 (lemma_pf ... trans\
 (lemma_pf ... def_i\
 (lemma_pf ... def_e\ 
 (def_pf form
    (Assoc\ pi F\ pi T\
      hastype (Assoc F T) form <<==
      pi X\ pi Y\
       (hastype X T ==>> hastype Y T ==>> hastype (F X Y) T))
    (F\T\ (forall T X\ forall T Y\ forall T Z\
     (eq T (F X (F Y Z)) (F (F X Y) Z))))
    (assoc\
 (lemma_pf
   (Assoc_inst\ pi F\ pi T\ pi A\ pi B\ pi C\ pi Q\
     proves (Assoc_inst F Q) (eq T (F A (F B C)) (F (F A B) C)) <<==
       hastype A T, hastype B T, hastype C T,
       pi X\ pi Y\ (hastype X T ==>> hastype Y T ==>> hastype (F X Y) T),
       proves Q (assoc F T))
   (F\Q\ 
    (elam T\ elam A\ elam B\ elam C\
     (extract (eq T (F A (F B C)) (F (F A B) C))
      (forall_e T (Z\ (eq T (F A (F B Z)) (F (F A B) Z)))
       (forall_e T (Y\ (forall T Z\ (eq T (F A (F Y Z)) (F (F A Y) Z))))
        (forall_e T (X\ (forall T Y\ (forall T Z\
                     (eq T (F X (F Y Z)) (F (F X Y) Z)))))
         (def_e form (assoc F T) 
          (forall T X\ forall T Y\ forall T Z\
           (eq T (F X (F Y Z)) (F (F X Y) Z))) (x\x) def Q) A) B) C))))
   (assoc_inst\
 (imp_i q1\ (forall_i a\ (imp_e (assoc f t)
  (imp_i q2\ (trans (f (f a a) (f a a))
              (assoc_inst f q2) (assoc_inst f q2)))
  (def_i form (assoc f t)
   (forall t a\ forall t b\ forall t c\
        (eq t (f a (f b c)) (f (f a b) c))) (x\x) def q1))))))))))))
 ((forall t a\ forall t b\ forall t c\
   eq t (f a (f b c)) (f (f a b) c)) imp
  (forall t a\ eq t (f a (f a (f a a))) (f (f (f a a) a) a))))
\end{verbatim}
\captionx{Theorem}{$(\forall a,b,c\; f a (f b c)=f(f a b) c) \supset
\forall a\; f a (f a (f a a)) = f (f (f a a) a) a$.}
\label{thm:def}
\end{figure}
The proof (the first argument to the \cd{proves} predicate) contains a
series of four lemmas which we have already seen,
followed by the definition of associativity,
followed by a fifth lemma about associativity (\cd{assoc\_inst}),
followed by the main body of the proof.  The \cd{def\_i} lemma is used
in the main body of the proof.  In general, proof checking using the
\cd{def\_i} lemma means that the proof being checked must match the term
\cd{(Def\_i T Name B P Q1 Q2)}, which is the first argument (the proof
term) of the head of the \cd{proves} clause implementing the
\cd{def\_i} lemma in Program~\ref{defs-fig}.  This match determines
the terms matching \cd{P} and \cd{Name}.  The formula being proved
must be a formula that matches the term \cd{(P Name)}, which is the
second argument of the head of the \cd{proves} clause implementing the
\cd{def\_i} lemma in Program~\ref{defs-fig}.
Here \cd{Name} is not always simply a 
variable name, but is actually the definition name applied to all of
its arguments to form a term of type \cd{tm}.  In our example,
\cd{assoc} has type
\begin{verbatim}
(tm -> tm -> tm) -> tp -> tm.
\end{verbatim}
At the point that proof checking of the body of the proof uses the
\cd{def\_i} lemma, the formula to be checked is \cd{(assoc f t)}.  The
term that corresponds to \cd{(P Name)} in this example is
\verb|(x\x)(assoc f t)|, which matches this formula.  Proof checking
proceeds by finding a proof of the goal of the form
\begin{verbatim}
(proves Q1 (eq form (assoc f t) B))
\end{verbatim}
which is proved simply by matching with the \lprolog{} equality
assumption added when the \cd{assoc} definition was processed by the
\cd{proves} clause for \cd{def\_pf}.  Next, the two typechecking
subgoals of the \cd{def\_i} clause are solved.  Solving the first,
\cd{(hastype (assoc f t) form)}, requires using the \lprolog{} type
inference assumption which was also added when the \cd{assoc}
definition was processed by the \cd{proves} clause for
\cd{def\_pf}.  Finally, the rest of the proof, is checked
via the subgoal of the form \cd{(proves Q2 (P B))}, where the formula
to be checked has the definition name replaced by its body.

The \cd{def\_e} lemma is used in the proof of the \cd{assoc\_inst}
lemma.  Its use in proof checking is similar to \cd{def\_i}.  The main
difference is that the formula to be checked must match the term
\cd{(P B)}, i.e., the formula contains an instance or instances of the
body of the definition, and in the subgoal to be checked, the body of
the definition is replaced with the name of the definition.

As another example of definitions, Program~\ref{def:and} shows the
definition of logical conjunction for the object logic using the
\cd{def\_pf} proof constructor.
\begin{figure}
\begin{verbatim}
 (def_pf form
    (And\ pi A\ pi B\
      hastype (And A B) form <<==
      hastype A form, hastype B form)
    (A\B\ (forall form C\ ((A imp B imp C) imp C)))
    (and\ ...
\end{verbatim}
\captionx{Program}{Definition of logical conjunction in the object logic.}
\label{def:and}
\end{figure}
Other connectives such as disjunction, negation, and existential
quantification can also be defined, and the rules for introduction and
elimination of these connectives can be proved as lemmas.

\subsection{An alternative implementation of definitions}
\label{defsSimple}

The new primitives and clauses in Program~\ref{definition-fig} provide
a convenient way of incorporating definitions, but actually are not
needed at all.  Instead, for each new definition, it is possible to
introduce a special lemma to handle that definition.  These special
lemmas are quite complex and we do not want to require the user to
come up with them.  For illustration, Theorem~\ref{thm:altdef} shows
the part of the proof that replaces the \cd{def\_pf} node in
Theorem~\ref{thm:def}.
\begin{figure}
\begin{verbatim}
...
 (lemma_pf ... def_e\ 
 (lemma_pf
   (Define_Assoc\ pi Q\ pi B\
     proves (Define_Assoc Q) B <<==
       pi d\ pi q\ 
          (pi F\ pi T\
               (proves q (eq form (d F T)
                              (forall T X\ forall T Y\ forall T Z\
                               (eq T (F X (F Y Z)) (F (F X Y) Z))))))
           ==>>
           (pi F\ pi T\ hastype (d F T) form <<==
                pi X\ pi Y\ hastype X T ==>> hastype Y T ==>>
                               hastype (F X Y) T)
           ==>> proves (Q d q) B)
   (Q\ (Q (F\T\ (forall T X\ forall T Y\ forall T Z\
                (eq T (F X (F Y Z)) (F (F X Y) Z)))) refl))
   (define_assoc\
 (define_assoc
   (assoc\q\
 (lemma_pf
   (Assoc_inst\ ...
\end{verbatim}
\captionx{Theorem}{Alternate proof of Theorem~\ref{thm:def}.}
\label{thm:altdef}
\end{figure}
This part of the proof includes the specialized lemma, called
\cd{Define\_Assoc}, and shows that it is used immediately after it is
defined. The bound variable \cd{assoc} represents the name for the new
definition, and the bound variable \cd{q} represents a proof of
equality between the definition name and its body.  The new proof
contains no use of the \cd{def\_pf} or \cd{def} proof constructors.
Occurrences of \cd{def} in Theorem~\ref{thm:def} are replaced with
\cd{q}.  This change, although not shown in Theorem~\ref{thm:altdef},
is the only other change required to obtain the complete alternate
proof.  We omit a detailed explanation of the \cd{Define\_Assoc}
lemma and simply note that it is fairly complex and increases the size
of this example proof.  Also, this lemma is similar in structure to
the simpler \cd{define} lemma described below in Section~\ref{defsAtomic}

Additional programming can make this alternative way of incorporating
definitions easier to use.  In particular, it is possible to write a
program to transform proofs that use \cd{def\_pf} and \cd{def} to
proofs that use only specialized lemmas such as the one in
Theorem~\ref{thm:altdef}.  Such a program would allow us to remove
Program~\ref{definition-fig} from the TCB.

\subsection{Handling atomic definitions}
\label{defsAtomic}

For the special class of defined terms that have meta-level type
\cd{tm}, which we call \emph{atomic} definitions, it is easy to
eliminate the need for \cd{def\_pf} and
\cd{def} because it is possible to include one new general lemma that
replaces them.
For example, we can express associativity of integers as the following
term
\begin{verbatim}
lam F\ forall intty X\ forall intty Y\ forall intty Z\
   eq intty (app intty (app intty F X) (app intty (app intty F Y) Z))
        (app intty (app F intty (app intty (app intty F X) Y)) Z)))
\end{verbatim}
where \cd{F} has meta-type \cd{tm} and object type \cd{(intty arrow
intty arrow intty)}, and the \cd{app} constructor is used to apply
\cd{F} to its arguments.  If we specialize Theorem~\ref{thm:def} to
integers, Theorem~\ref{thm:newdef} shows the part of the proof of this
new theorem that replaces what is shown in Theorem~\ref{thm:altdef}.
\begin{figure}
\begin{verbatim}
...
 (lemma_pf ... def_e\ 
 (lemma_pf
   (Define\ pi T\ pi F\ pi Q\ pi B\
     proves (Define T F Q) B <<==
      hastype F T,
      pi d\ pi q\ (hastype d T ==>>
                    proves q (eq T d F) ==>> proves (Q d q) B))
   (T\F\P\ (P F refl))
   (define\
 (define ((intty arrow intty arrow intty) arrow form)
   (lam F\ forall intty X\ forall intty Y\ forall intty Z\
       eq intty (app intty (app intty F X) (app intty (app intty F Y) Z))
            (app intty (app intty F (app intty (app intty F X) Y)) Z)))
   (assoc\q\
 (lemma_pf
   (Assoc_inst\ ...
\end{verbatim}
\captionx{Theorem}{Alternate proof of Theorem~\ref{thm:def}
specialized to integers.}
\label{thm:newdef}
\end{figure}
The parts of the proof not shown are similar to Theorems~\ref{thm:def}
and~\ref{thm:altdef}, but modified to use the new type of the bound
variable \cd{f}, which has the same type as the bound \cd{F} in the
definition.

In general, to check a proof using the \cd{define} lemma, which has
the following form
\begin{verbatim}
(define T Term (Name\ EqProof\ (RestProof Name EqProof)))
\end{verbatim}
the system interprets the ``\cd{pi d}'' within the \cd{define} lemma
to create a new atom \cd{d} to stand for the \cd{Name}.  The new atom
\cd{q} is also introduced to stand for a proof that the name is equal
to the body of the definition, and \cd{(proves q (eq T d Term))} is
added to the clause database.  Finally, $\beta$-conversion substitutes
\cd{d} for \cd{Name} and \cd{q} for \cd{EqProof} within
\cd{RestProof} and the resulting proof is checked.

In proof checking the new proof, instead of subproofs of the form
\begin{verbatim}
(proves def (eq form (assoc f t) B))
\end{verbatim}
that would be generated by proof checking Thereom~\ref{thm:def}, or
subproofs of the form
\begin{verbatim}
(proves q (eq form (assoc f t) B))
\end{verbatim}
that would be generated by proof checking Thereom~\ref{thm:altdef}, in
Theorem~\ref{thm:newdef} we have subproofs of the form
\begin{verbatim}
(proves q (eq ((intty arrow intty arrow intty) arrow form) assoc B))
\end{verbatim}
where \cd{q} here is the name of the proof term introduced inside the
\cd{define} proof node.

In general, having a single \cd{define} lemma that can be used by all
atomic definitions is simpler, but the atomic forms of definitions are
larger and harder to read.  In the case of \cd{assoc}, the atomic
version is three lines, while the original version is one line long.
In our previous work~\cite{AppelFelty:ICLP99}, having to choose
between the version of
\cd{assoc} that used \cd{app} and the one that didn't was not an
issue, since there were no \cd{app} and \cd{lam} constructors.  Instead
application and abstraction were encoded directly using application
and abstraction at the meta-level.  Also, there was no reason to
include a separate \cd{def\_pf} proof constructor; the \cd{define}
lemma was sufficient for introducing all definitions.  Although this
allowed a simpler version of definitions, we were unable to allow
polymorphism in definitions, which is desirable in definitions for the
same reason it is desirable in lemmas.  Our previous encoding
also did not allow definitions for
object-level types.  For example, in the domain of proof-carrying
code, we have declarations like this one
\begin{verbatim}
hastype has_mltype
        ((exp arrow form) arrow (exp arrow exp) arrow exp arrow
         ((exp arrow form) arrow (exp arrow exp) arrow exp arrow
          form) arrow form.
\end{verbatim}
Types like this arise because we encode types of the programming
language we are reasoning about (in this case ML) as predicates which
themselves take predicates as arguments.  In our new version, it is
possible to handle definitions at meta-type \cd{tp}; we would need a
new proof constructor and a new proof checking clause similar to the
one for the \cd{def\_pf} proof constructor in
Program~\ref{definition-fig}.  Adding type definitions would also
require adding reasoning about equality of types into our typechecking
clauses.

\section{Programming with lemmas and definitions}
\label{sec:prog}

The lemma and definition mechanisms provide ways to store lemmas and
definitions inside proofs.  Packaging proofs in this way makes it
straightforward to communicate proofs, and keeps the proof checking
machinery (the TCB) simple, which is important for our proof-carrying
code application.  Thus far, all the \lprolog{} code in
Programs~\ref{core-types}, \ref{core-rules}, \ref{lemmafig},
\ref{cinterp}, \ref{validclause}, \ref{teyjuslemmafig}, \ref{elam},
and \ref{definition-fig} is inside the TCB.  A good environment for
building proofs is also essential, and this part of the code can be
outside the TCB.  We don't have to be as careful because we know that
any proofs we build in our theorem proving environment have to be
checkable by the proof checking code presented so far.

As we build a library of lemmas and definitions, we clearly don't want
to store every lemma and definition inside every proof that uses them.
Instead, for lemmas that have general applicability like \cd{symm}, we
would like to store them each once and allow them to be used in other
proofs as needed.  To do so, we provide predicates for stating each
definition and lemma.  To use these predicates, we must introduce new
constants for definition and lemma names.  Program~\ref{prog:store}
contains the declarations of these new predicates, and two examples
which use them.
\begin{figure}
\begin{verbatim}
type  def_lemma       A -> (A -> o) -> A -> o.
type  def_definition  tp -> A -> (A -> o) -> A -> o.

type symm             pf -> pf.
def_lemma symm
  (Symm\ pi T\ pi A\ pi B\ pi P\
    proves (Symm P) (eq T A B) <<== 
        hastype A T, hastype B T, proves P (eq T B A))
  (P\ elam A\ elam B\ elam T\
    (extract (eq T A B) (congr T B A (eq T A) P refl))).

type  assoc           (tm -> tm -> tm) -> tp -> tm.
def_definition form assoc
  (Assoc\ pi F\ pi T\
    hastype (Assoc F T) form <<==
    pi X\ pi Y\
     (hastype X T ==>> hastype Y T ==>> hastype (F X Y) T))
  (F\T\ (forall T X\ forall T Y\ forall T Z\
   (eq T (F X (F Y Z)) (F (F X Y) Z)))).
\end{verbatim}
\captionx{Program}{Storing lemmas and definitions.}
\label{prog:store}
\end{figure}
\lprolog{}'s polymorphism is used in these predicates.
The first argument to \cd{def\_lemma} is the lemma name, and
the next two arguments correspond to the \cd{Inference} and
\cd{LemmaProof} arguments to the \cd{lemma\_pf} constructor.  The arguments
to \cd{def\_definition} are the definition name (the second argument)
and arguments that correspond to the first three arguments of the
\cd{def\_pf} constructor (arguments 1, 3, and 4 here).

Then we can write programs to manipulate these lemmas and definitions
in various ways.  For example, if we want to package a proof as a
single term with all the definitions and lemmas it depends on inside
it, we must write a program to do so.  The resulting proof should not
contain any constants like \cd{symm} and \cd{assoc}; instead lemma and
definition names must be bound variables inside occurrences of the
\cd{lemma\_pf} and \cd{def\_pf} proof constructors.  We do not present
the ``packaging'' program here, but instead present a simpler program
that illustrates some of the programming techniques required for
manipulating lemmas and definitions stored in this way.
Program~\ref{prog:check} contains a program for checking a proof.  It
doesn't check the lemmas that the proof depends on, but could be
easily modified to do so.
\begin{figure}
\begin{verbatim}
type  done_def        A -> o.
type  done_lemma      A -> o.
type  check_lem       A -> o.
type  check_lem_aux   B -> A -> (A -> o) -> A -> o.

check_lem Name :-
  def_definition T DName Inference Def,
  not (done_def DName), !,
  def_to_eqclause T DName Def EqClause,
  done_def DName => cl (Inference DName) => cl EqClause => check_lem Name.

check_lem Name :-
  def_lemma LName Inference LemmaProof,
  check_lem_aux Name LName Inference LemmaProof.

check_lem_aux Name Name Inference LemmaProof :- !,
  pi name\ (valid_clause (Inference name)),
  (Inference LemmaProof).

check_lem_aux Name LName Inference LemmaProof :- !,
  not (done_lemma LName), !,
  done_lemma LName => cl (Inference LName) => check_lem Name.
\end{verbatim}
\captionx{Program}{Checking a proof which uses stored lemmas and
definitions.}
\label{prog:check}
\end{figure}
The trick of using Prolog cut (\cd{!}) along with the predicates
\cd{done\_def} and \cd{done\_lemma} allows us to process a list of
clauses in the order they
appear in the database.  The first clause for \cd{check\_lem} looks
for the next definition and each time it finds a new one, it adds the
corresponding typechecking clause and equality clause.  The second
\cd{check\_lem} clause is used once all definitions have been added.
It finds the next lemma and uses \cd{check\_lem\_aux} to see if the next
lemma is the one that should be checked.  If so, the proof is checked;
if not, the proof checking clause for the lemma is added to the
database and \cd{check\_lem} is called to process the next lemma.

\section{Encoding the core logic in Twelf}
\label{sec:twelf}

The {\em Logical Framework} (LF)~\cite{Harper89} is another example of
a metalanguage in which it is possible to encode a wide variety of
logics.  The Twelf system~\cite{Pfenning99b} is an implementation of
LF which provides logic programming capabilities, many of which are
similar to \lprolog{}.  In this section, we compare the encoding of
our core logic in \lprolog{} to a corresponding encoding in Twelf,
discuss lemmas and definitions in Twelf, and compare the programming
environments of these two languages.

\subsection{The core logic in Twelf}
\label{encode}

LF is a $\lambda$-calculus with dependent types.  A dependent type in
LF has the structure $\{x:A\}B$ where $A$ and $B$ are types and $x$ is
a variable of type $A$ bound in this expression.  The type $B$ may
contain occurrences of $x$.  This structure represents a ``functional
type.''  If $f$ is a function of this type, and $N$ is a term of type
$A$, then $fN$ ($f$ applied to $N$) has the type $[N/x]B$, which
represents the type $B$ where all occurrences of $x$ are replaced by
$N$.  Thus the argument type is $A$ and the result type {\em depends}
on the value input to the function.  If $x$ doesn't occur in $B$, this
type is often abbreviated using the usual type arrow: $A\rightarrow
B$.

The extra expressiveness of dependent types allows object-level types
to be expressed more directly, eliminating the need for any
typechecking clauses like the \cd{hastype} clauses of
Program~\ref{core-rules}.  The Twelf constructor declarations in
Program~\ref{core-twelf} illustrate the use of dependent types for
encoding our object logic.
\begin{figure}
\begin{verbatim}
tp     : type.
tm     : tp -> type.

form   : tp.
pf     : tm form -> type.

intty  : tp.
arrow  : tp -> tp -> tp.                     %infix right 14 arrow.
pair   : tp -> tp -> tp.

eq     : tm T -> tm T -> tm form.
imp    : tm form -> tm form -> tm form.      %infix right 10 imp.
forall : (tm T -> tm form) -> tm form.
false  : tm form.

lam    : (tm T1 -> tm T2) -> tm (T1 arrow T2).
app    : tm (T1 arrow T2) -> tm T1 -> tm T2.
mkpair : tm T1 -> tm T2 -> tm (pair T1 T2).
fst    : tm (pair T1 T2) -> tm T1.
snd    : tm (pair T1 T2) -> tm T2.

refl     : pf (eq X X).
beta     : pf (eq (app (lam F) X) (F X)).
fstpair  : pf (eq (fst (mkpair X Y)) X).
sndpair  : pf (eq (snd (mkpair X Y)) Y).
surjpair : pf (eq (mkpair (fst Z) (snd Z)) Z).
congr    : {H: tm T -> tm form}
             pf (eq X Z) -> pf (H Z) -> pf (H X).
imp_i    : (pf A -> pf B) -> pf (A imp B).
imp_e    : pf (A imp B) -> pf A -> pf B.
forall_i : ({y:tm T}pf (A y)) -> pf (forall A).
forall_e : pf (forall A) -> {y:tm T}pf (A y).
\end{verbatim}
\captionx{Program}{Core logic in Twelf.}
\label{core-twelf}
\end{figure}
Felty and Miller~\cite{FeltyMillerCADE90} show how to transform an LF
object logic into an encoding in a higher-order logic which is a
sublogic of the one implemented by \lprolog.  The discussion in this
section is informal, but in Appendix~\ref{app:compare}, we use this
transformation to provide a formal basis for comparing our two
encodings.

Although typechecking clauses are not needed here, the proof checking
operation is more complicated in Twelf since it requires type
reconstruction for dependent types.

\subsection{Lemmas and definitions in Twelf}

Twelf has its own built-in definition mechanism, which can be used for
both lemmas and definitions in the object logic.
Program~\ref{twelf-def} contains a Twelf version of the definition of
\cd{assoc} and the \cd{symm} lemma.
\begin{figure}
\begin{verbatim}
%abbrev
assoc : (tm T -> tm T -> tm T) -> tm form =
  [f:(tm T -> tm T -> tm T)]
   (forall [a:tm T] forall [b:tm T] forall [c:tm T]
    (eq (f a (f b c)) (f (f a b) c))).

symm: pf (eq X Y) -> pf (eq Y X) =
 [q:pf (eq X Y)] (congr ([z:tm T] (eq Y z)) q refl).
\end{verbatim}
\captionx{Program}{Example lemmas and definitions in Twelf.}
\label{twelf-def}
\end{figure}
The \cd{abbrev} directive is required in some definitions for
technical reasons, which we do not describe here.  There are three
parts to a definition: a constant naming the definition, its type, and
its body (an LF term).  A lemma is similar and contains its name, the
formula representing the statement of the lemma (which is a type in
LF), and the proof (an LF term).

In Twelf, a proof is simply a series of declarations and definitions,
where the last one is the statement and proof of the main theorem.
This proof possibly depends on the lemmas and definitions that come
before it.  Each definition in the sequence has the form mentioned
above: a name, a type, and the term which the name abbreviates when it
appears in subsequent declarations.  The declarations defining the
logical constants and primitive inference rules shown in
Program~\ref{core-twelf} (which each have a type but no defining term)
are at the beginning of the sequence.  In Twelf, we cannot package up
a lemma and its proof, or a definition and its body, along with the
rest of the proof, in the same way we did in \lprolog.  The reason for
this is that we cannot introduce a \cd{lemma\_pf} or \cd{def\_pf}
constructor because they require polymorphism at the meta-level, which
Twelf does not have.

In our \lprolog{} version, we discussed naming each lemma and
definition, including one copy of each in a library, and using it
whenever needed.  We then presented a program which was able to check
the proof of a theorem, assuming that lemmas and definitions were
organized in this way.  In Twelf, we don't need a special program for
checking proofs of lemmas.  One of the central meta-operations of
Twelf is to read in a series of declarations and definitions, and
check each one as it is encountered.  Proofs are fully checked by this
operation.

In Twelf, other kinds of operations on proofs are limited.  Many proof
transformations that we can implement in \lprolog{} are not
programmable in Twelf either because they require polymorphism or
because they require manipulation of meta-level formulas.
Manipulation of meta-level formulas is not possible in Twelf because
it requires quantification over such formulas (i.e., quantification
over types containing \cd{type}), which is not allowed.

\section{Other issues}

Although we have focussed on the lemma and definition mechanisms
in \lprolog{} and Twelf,
other aspects of the metalanguage are also relevant to our needs for
proof generation and checking.

\subsection{Arithmetic}
\label{arith}

For our application, proof-carrying code, we wish to prove theorems
about machine instructions that add, subtract, and multiply; and about
load/store instructions that add offsets to registers.  Therefore we
require some rudimentary integer arithmetic in our logic.

Some logical frameworks have powerful arithmetic primitives, such as
the ability to solve linear programs \cite{necula98:phd} or to handle
general arithmetic constraints \cite{JaffarLassez87}.
For example, Twelf provides a complete theory of the rationals,
implemented using linear programming \cite{virga:twelfx}.
On the one hand, linear programming is a powerful and
general proof technique, although it can increase the
complexity of the TCB.  On the other hand,
synthesizing arithmetic from scratch is not easy.  We have also
experimented with arithmetic in \lprolog{} where we use the \cd{is}
predicate to provide some automatic simplifications.

\subsection{Representing proof terms}

Parameterizable data structures with higher-order unification modulo
$\beta$-equivalence provide an expressive way of representing
formulas, predicates, and proofs.  We make heavy use of higher-order
data structures with both direct sharing and sharing modulo
$\beta$-reduction.  The implementation of the metalanguage must
preserve this sharing; otherwise our proof terms will blow up in size.

Any logic programming system is likely to implement sharing of terms
obtained by copying multiple pointers to the same subterm.  In Terzo,
this can be seen as the implementation of a reduction algorithm
described by Wadsworth \cite{wadsworth71}.  But we require even more
sharing.  The similar terms obtained by applying a $\lambda$-term to
different arguments should retain as much sharing as possible.
Therefore some intelligent implementation of higher-order terms within
the metalanguage---such as Teyjus's use of explicit
substitutions~\cite{NadathurLFP90,NadathurTCS98}---seems essential.

\subsection{Programming the prover}

In this paper, we have concentrated on an encoding of the logic used
for proof checking, and discussed some operations on proofs.  But of
course, we will also need to construct proofs.  For the proof-carrying
code application, we need an automatic theorem prover to prove the
safety of programs.  For implementing this prover, we have found that
the Prolog-style control primitives (such as the cut (\cd{!}) operator
and the \cd{is} predicate), which are also available in \lprolog{},
are quite important.  \lprolog{} also provides an environment for
implementing tactic-style interactive provers \cite{felty93}.  This
kind of prover is useful for proving the lemmas that are used by the
automatic prover.

Twelf does not have many control primitives; in fact, implementation
of control primitives does not fit well into the Twelf system design.
We have begun to experiment with an operator in Twelf similar to
Prolog cut, to see if it will allow us to implement the automatic
prover in the same way as in \lprolog.  There is also no support for
building interactive provers in Twelf, so proofs of lemmas used by the
automatic prover must be constructed by hand.

\section{Conclusion}

The logical frameworks discussed in this paper are promising vehicles
for proof-carrying code, or in general where it is desired to keep the
proof checker as small and simple as possible.  We have proposed a
representation for lemmas and definitions that should help keep proofs
small and well-structured, and each of these
frameworks has features that are useful in implementing, or
implementing efficiently, our machinery.

We have found the conciseness of the encoding in Twelf to be
particularly convenient, and because of that, we have used Twelf for
extensive proof development in our proof-carrying code application.
As programming with proofs becomes more important in the next phases
of our system, \lprolog{} will have more advantages.  We are currently
investigating ways to combine the use of the two metalanguages.  The
translation discussed in Appendix~\ref{app:compare} will serve as the
foundation for this combination.

\appendix
\section{A full interpreter for proof checking}
\label{dynamic}

To write a full interpreter, we extend Program ~\ref{cinterp} in
Section~\ref{lemmasTeyjus} by introducing a new type \cd{goal} and
connectives which build terms of this type.  In particular, we now
give \cd{<<==} and \cd{==>>} the type
\cd{goal -> goal -> goal}.  We also introduce a new constant
\verb|^^| for conjunction having the same type as the implication
constructors.  Finally, we introduce \cd{all} for universal
quantification having type \cd{(A -> goal) -> goal}.  In
addition, we change the type of \cd{backchain} to
\cd{goal -> goal -> o}, and modify the clauses for the \emph{comma} and
\cd{pi} to use the new constants.  In the \cd{backchain} clauses for
\cd{<<==} and \cd{==>>} in Program ~\ref{cinterp}, the goal \cd{G1} which
appears as an argument inside the head of the clause also appears as a
goal in the body of the clause.  In the full interpreter, we cannot do
this.  \cd{G1} no longer has type
\cd{o}; it has type \cd{goal} and is constructed using the new
connectives.  Instead, we replace \cd{G1} with \cd{(solveg G1)} and
implement the \cd{solveg} predicate to handle the solving of goals.
The new code for \cd{solveg} and the modified code for
\cd{backchain} is in Program~\ref{interp}.
\begin{figure}
\begin{verbatim}
kind  goal            type.

type  ==>>            goal -> goal -> goal.     infixr  ==>>    4.
type  <<==            goal -> goal -> goal.     infixl  <<==    0.
type  ^^              goal -> goal -> goal.     infixl  ^^      3.
type  all             (A -> goal) -> goal.

type  cl              goal -> o.
type  backchain       goal -> goal -> o.
type  solveg          goal -> o.

type  proves          pf -> form -> goal.
type  assume          form -> goal.
type  valid_clause    goal -> goal.

solveg (all G) :- pi x\ (solveg (G x)).
solveg (G1 ^^ G2) :- solveg G1, solveg G2.
solveg (D ==>> G) :- (cl D) => solveg G.
solveg (G <<== D) :- (cl D) => solveg G.
solveg G :- cl D, backchain G D.

backchain G G.
backchain G (all D) :- backchain G (D X).
backchain G (A ^^ B) :- backchain G A; backchain G B.
backchain G (H <<== G1) :- backchain G H, solveg G1.
backchain G (G1 ==>> H) :- backchain G H, solveg G1.
\end{verbatim}
\captionx{Program}{A full interpreter.}
\label{interp}
\end{figure}
In order to use this interpreter to solve goals of the form
\cd{(proves P A)}, the \cd{proves} predicate must be a constructor for
terms of type \cd{goal}, and the meta-level goal presented to \lprolog{}
must have the form \cd{(solveg (proves P A))}.  Similarly, inference
rules must also be represented as objects of type \cd{goal} and
wrapped inside \cd{cl} to form \lprolog{} clauses.  Several examples
of clauses for inference rules are given in
Program~\ref{inf-interp} to illustrate.
\begin{figure}
\begin{verbatim}
cl (proves Q A <<== assump (proves Q A)).
cl (proves (imp_i Q) (A imp B) <<==
      all p\ (assump (proves p A) ==>> proves (Q p) B)).
cl (proves (forall_i Q) (forall T A) <<==
      all y\ (hastype y T ==>> proves (Q y) (A y))).
cl (proves (lemma_pf Inference LemmaProof RestProof) C <<==
  all Name\ 
   (valid_clause (Inference Name) ^^
    Inference LemmaProof ^^
    (Inference Name) ==>> (proves (RestProof Name) C))).
\end{verbatim}
\captionx{Program}{Clauses used by the full interpreter.}
\label{inf-interp}
\end{figure}
The last clause is the new clause for handling lemmas in this setting.
Note that in this version, \cd{valid\_clause} constructs objects of
type \cd{goal}; thus all the clauses for \cd{valid\_clause} must also
be wrapped in \cd{cl}.

\section{Comparison of the core logic in Twelf and \lprolog}
\label{app:compare}

As stated, the transformation in Felty and
Miller~\cite{FeltyMillerCADE90} can provide a formal basis for
comparing our two encodings. In order to perform this transformation,
we must consider a ``full'' LF encoding, which does not take advantage
of the abbreviations that Twelf allows.  Just as the full LF encoding
can be improved by using Twelf's abbreviations, the \lprolog{} program
that results from the transformation can be improved by making several
optimizations.  We discuss how the encoding presented in
Programs~\ref{core-types} and~\ref{core-rules} can be viewed as the
application of the transformation, followed by performing several such
optimizations.

In both \lprolog{} and Twelf, all tokens in a clause or declaration
beginning with
uppercase letters are implicitly bound by universal quantifiers at the
outermost level.  In Twelf, this implicit quantification is important
for providing an encoding of the object logic that is readable and
usable.  To see why, consider the \texttt{surjpair} rule, which uses
the \texttt{mkpair}, \texttt{fst}, and \texttt{snd} constants.  We can
make the outermost quantification explicit in Twelf, resulting in the
declarations:
\begin{verbatim}
mkpair : {T1:tp}{T2:tp}tm T1 -> tm T2 -> tm (pair T1 T2).
fst    : {T1:tp}{T2:tp}tm (pair T1 T2) -> tm T1.
snd    : {T1:tp}{T2:tp}tm (pair T1 T2) -> tm T2.
surjpair :
 {T1:tp}{T2:tp}{Z:tm (pair T1 T2)}
 pf (eq (pair T1 T2) (mkpair T1 T2 (fst T1 T2 Z) (snd T1 T2 Z)) Z).
\end{verbatim}
This version of \texttt{surjpair} is quite a bit bigger than the one
in Program~\ref{core-twelf}.  Explicitly including \texttt{T1} and
\texttt{T2} means that \texttt{mkpair}, \texttt{fst}, and \texttt{snd}
each take two extra type arguments, while \texttt{surjpair} takes
three.  Terms containing these constants must then take extra arguments
which in this example causes redundancy in the type of
\texttt{surjpair} because the same types appear many times.
Implicit quantifiers make the encoding easier to read and work with.
In fact, in the version we used in our experiments, 
the fact that \cd{app} could be represented as a binary constructor
without loss of information allowed us to replace the 
\texttt{app} constant with an infix symbol, resulting in
encoded terms that were syntactically even closer to the terms
they represented.  We cannot make \texttt{app} in the \lprolog{} encoding
infix because it takes three arguments.  (We discuss why it must take
three arguments below.)

The explicit quantifiers that we have left out in
Program~\ref{core-twelf} are those that Twelf can easily reconstruct.
Because of this reconstruction, however, a Twelf
typechecker (proof checker) has to work harder than it would if we
used an explicit version.  These encodings illustrate a
tradeoff we encounter in proof and term size versus complexity of the proof
checker.  Reducing the proof size forces the checker (the TCB)
to become more complex.

When considering the formal transformation, we start from a modified
version of Program~\ref{core-twelf} that makes all quantifiers
explicit.  To illustrate, we apply the transformation to all of the
declarations in the Twelf encoding except for the constants and
inference rules for pairing.  Applying the transformation to these
declarations, we get the \lprolog{} type declarations and clauses in
Programs~\ref{trans-types} and~\ref{trans-rules}.
\begin{figure}
\begin{verbatim}
kind  ltp             type.
kind  ltm             type.

type  ltype           ltp -> o.
type  hasltype        ltm -> ltp -> o.
type  well_typed      ltm -> ltp -> o.

type  tp              ltp.
type  tm              ltm -> ltp.

type  form            ltm.
type  pf              ltm -> ltp.

type  intty           ltm.
type  arrow           ltm -> ltm -> ltm.        infixr  arrow   8.

type  lam             ltm -> ltm -> (ltm -> ltm) -> ltm.
type  app             ltm -> ltm -> ltm -> ltm -> ltm.
type  eq              ltm -> ltm -> ltm -> ltm.
type  imp             ltm -> ltm -> ltm.        infixr  imp     7.
type  forall          ltm -> (ltm -> ltm) -> ltm.
type  false           ltm.

type  refl            ltm -> ltm -> ltm.
type  beta            ltm -> ltm -> (ltm -> ltm) -> ltm -> ltm.
type  congr           ltm -> ltm -> ltm -> (ltm -> ltm) ->
                         ltm -> ltm -> ltm.
type  imp_i           ltm -> ltm -> (ltm -> ltm) -> ltm.
type  imp_e           ltm -> ltm -> ltm -> ltm -> ltm.
type  forall_i        ltm -> (ltm -> ltm) -> (ltm -> ltm) -> ltm.
type  forall_e        ltm -> (ltm -> ltm) -> ltm -> ltm -> ltm.
\end{verbatim}
\captionx{Program}{Type declarations for transformation of Twelf to \lprolog.}
\label{trans-types}
\end{figure}
\begin{figure}
\begin{verbatim}
well_typed M A :- ltype A, hasltype M A.

ltype tp.
ltype (tm T) :- hasltype T tp.
ltype (pf A) :- hasltype A (tm form).

hasltype intty tp.
hasltype form tp.
hasltype (T1 arrow T2) tp :- hasltype T1 tp, hasltype T2 tp.

hasltype (lam T1 T2 F) (tm (T1 arrow T2)) :- hasltype T1 tp, hasltype T2 tp,
  pi x\ (hasltype x (tm T1) => hasltype (F x) (tm T2)).
hasltype (app T1 T2 F X) (tm T2) :- hasltype T1 tp, hasltype T2 tp,
  hasltype F (tm (T1 arrow T2)), hasltype X (tm T1).
hasltype (eq T X Y) (tm form) :-
  hasltype T tp, hasltype X (tm T), hasltype Y (tm T).
hasltype (A imp B) (tm form) :- hasltype A (tm form), hasltype B (tm form).
hasltype (forall T A) (tm form) :- hasltype T tp,
  pi x\ (hasltype x (tm T) => hasltype (A x) (tm form)).
hasltype false (tm form).

hasltype (refl T X) (pf (eq T X X)) :- hasltype T tp, hasltype X (tm T).
hasltype (beta T1 T2 F X) (pf (eq T2 (app T1 T2 (lam T1 T2 F) X) (F X))) :-
  hasltype T1 tp, hasltype T2 tp,
  pi x\ (hasltype x (tm T1) => hasltype (F x) (tm T2)).
hasltype (congr T X Z H P1 P2) (pf (H X)) :-
  hasltype T tp, hasltype X (tm T), hasltype Z (tm T),
  pi x\ (hasltype x (tm T) => hasltype (H x) (tm form)),
  hasltype P1 (pf (eq T X Z)), hasltype P2 (pf (H Z)).
hasltype (imp_i A B Q) (pf (A imp B)) :-
  hasltype A (tm form), hasltype B (tm form).
  pi p\ (hasltype p (pf A) => hasltype (Q p) (pf B)).
hasltype (imp_e A B Q1 Q2) (pf B) :- 
  hasltype A (tm form), hasltype B (tm form),
  hasltype Q1 (pf (A imp B)), hasltype Q2 (pf A).
hasltype (forall_i T A Q) (pf (forall T A)) :- hasltype T tp,
  pi y\ (hasltype y (tm T) => hasltype (A y) (tm form)),
  pi y\ (hasltype y (tm T) => hasltype (Q y) (pf (A y))).
hasltype (forall_e T A Q Y) (pf (A Y)) :- hasltype T tp,
  pi y\ (hasltype y (tm T) => hasltype (A y) (tm form)),
  hasltype Q (pf (forall T A)), hasltype Y (tm T).
\end{verbatim}
\captionx{Program}{Transformation of Twelf declarations to \lprolog{}
clauses.}
\label{trans-rules}
\end{figure}
Before discussing the details, it is already possible to see some of
the similarities between the Twelf and \lprolog\ encodings, and
between the \lprolog\ encoding resulting from the transformation and
the one in Programs~\ref{core-types} and~\ref{core-rules}.  For
example, in Twelf the full version of the \cd{congr} rule is
\begin{verbatim}
congr    : {T:tp}{X:tm T}{Z:tm T}{H:tm T -> tm form}
             pf (eq X Z) -> pf (H Z) -> pf (H X).
\end{verbatim}
The \texttt{congr} proof constructor takes 6 arguments (\cd{T},
\cd{X}, \cd{Z}, \cd{H}, and two subproofs).
In the \lprolog\ version of \cd{congr} in Programs~\ref{trans-types}
and~\ref{trans-rules}, \cd{congr} also takes 6 arguments (4 terms and
2 subproofs) though their types are different from the LF version.
Also, in our original
\lprolog\ encoding (Program~\ref{core-rules}), the \texttt{congr}
clause has 4 subgoals, while in the new one
(Program~\ref{trans-rules}) there are 6; it is easy to see the
correspondence between 4 of them in the two encodings.
Note that in the version in Program~\ref{core-rules},
two of them are typechecking subgoals and two are
proof checking subgoals.  In Twelf, typechecking and proof checking
are unified, so all subgoals in the Twelf version are Twelf
typechecking goals; in our example some of them check terms whose
types have the form \cd{(tm A)}, while others check terms whose types
have the form \cd{(pf A)}.

In LF, there are several kinds of assertions.  The two that are
important for the formal transformation are: ``$A$ is a type''
and ``term $M$ has type $A$''.  Two \lprolog{} types \cd{ltp} and
\cd{ltm} introduced in Program~\ref{trans-types} are used to encode LF
types and terms.  The \lprolog{} predicates \cd{ltype} and
\cd{hasltype} are introduced to express the two assertions,
respectively.  The first assertion is important for transforming the
three declarations in Program~\ref{core-twelf} that end in
``\cd{type.}''  They declare constants that are used to create LF
types, which correspond to \lprolog\ formulas (terms of type \cd{o}).
The second assertion is used for the rest.  In order for an assertion
of the second kind to hold, it must also be the case that $A$ is a
type.  For this reason, the \lprolog{} predicate \cd{well\_typed} is
included (Program~\ref{trans-types}) and has one clause
(Program~\ref{trans-rules}).  The declarations and clause discussed so
far are necessary no matter what Twelf encoding we begin with.  The
remaining declarations and clauses in Programs~\ref{trans-types}
and~\ref{trans-rules} are specific to our particular object logic.
For each Twelf declaration in Program~\ref{core-twelf} that we
consider, there is one type declaration in Program~\ref{trans-types}
and one clause in Program~\ref{trans-rules}.

The first change we make to the \lprolog{} code in
Programs~\ref{trans-types} and~\ref{trans-rules} to get closer to an
optimized version involves the \cd{well\_typed} clause.  Consider the
first subgoal of this clause, an \cd{ltype} subgoal.  Note that for
our particular encoding,
there are three clauses for the \cd{ltype} predicate.  They correspond to
the three kinds of objects in the encoding of the object logic: types,
terms, and proofs.  In solving an \cd{ltype} subgoal, at most one
clause will ever apply at any point depending on which of three forms
the argument has.  This observation permits us to replace
\cd{well\_typed} with the following three clauses which cover every case.
\begin{verbatim}
well_typed T tp :- ltype tp, hasltype T tp.
well_typed M (tm T) :- ltype (tm T), hasltype M (tm T).
well_typed M (pf A) :- ltype (pf A), hasltype M (pf A).
\end{verbatim}
In the first clause, we can eliminate the \cd{ltype} subgoal because
it is always provable.  In the second and third clauses, we can
replace the \cd{ltype} subgoal with the corresponding subgoal from the
body of the only \cd{ltype} clause that applies, to obtain the clauses
below.
\begin{verbatim}
well_typed T tp :- hasltype T tp.
well_typed M (tm T) :- hasltype T tp, hasltype M (tm T).
well_typed M (pf A) :- hasltype A (tm form), hasltype M (pf A).
\end{verbatim}
Now, we no longer have a need for the \cd{ltype} clauses and can
eliminate them.

Although \cd{hasltype} is sufficient for representing any LF assertion
of the form ``term $M$ has type $A$,'' in our encoding it is useful to
distinguish three ways in which it is used.  This fact leads to our
second modification of Programs~\ref{trans-types}
and~\ref{trans-rules}.  The second argument to \cd{hasltype} always
has one of the following forms: \cd{tp}, \cd{(tm T)}, or \cd{(pf A)}.
Using this fact, we replace \cd{hasltype} with three predicates:
\cd{istype}, \cd{hastype}, and \cd{proves}.  Since the second argument
to \cd{istype} always is \cd{tp}, we can eliminate this argument
altogether so that \cd{istype} has type \cd{ltm -> o}.
Program~\ref{trans-mod1} illustrates the modifications discussed so
far on a subset of the \cd{hasltype} clauses in
Program~\ref{trans-rules}, which include only those for \cd{arrow},
\cd{forall}, and \cd{forall\_i}.
\begin{figure}
\begin{verbatim}
kind  ltp             type.
kind  ltm             type.

type  istype          ltm -> o.
type  hastype         ltm -> ltp -> o.
type  proves          ltm -> ltp -> o.
type  well_typed      ltm -> ltp -> o.

type  tp              ltp.
type  tm              ltm -> ltp.
type  pf              ltm -> ltp.

type  arrow           ltm -> ltm -> ltm.        infixr  arrow   8.
type  forall          ltm -> (ltm -> ltm) -> ltm.
type  forall_i        ltm -> (ltm -> ltm) -> (ltm -> ltm) -> ltm.

well_typed T tp :- istype T.
well_typed M (tm T) :- istype T, hastype M (tm T).
well_typed M (pf A) :- hastype A (tm form), proves M (pf A).

istype (T1 arrow T2) :- istype T1, istype T2.

hastype (forall T A) (tm form) :- istype T,
  pi x\ (hastype x (tm T) => hastype (A x) (tm form)).

proves (forall_i T A Q) (pf (forall T A)) :- istype T,
  pi y\ (hastype y (tm T) => hastype (A y) (tm form)),
  pi y\ (hastype y (tm T) => proves (Q y) (pf (A y))).
\end{verbatim}
\captionx{Program}{Modification of selected \lprolog{} declarations
and clauses from Programs~\ref{trans-types} and~\ref{trans-rules}.}
\label{trans-mod1}
\end{figure}

Looking back at Program~\ref{core-twelf}, note the types of the four
constants that are used to construct terms of type \cd{tp}.  There are
no dependent types here; they are all simple types, which could be
transformed directly to \lprolog{} types.  This fact leads to our
third modification.  Instead of transforming all Twelf terms and types
to \lprolog{} terms as is done by the transformation, we transform
types with no dependencies directly to \lprolog{} types, thus allowing
the \lprolog{} typechecker to do more typechecking work automatically.
This direct transformation gives us the \lprolog{} declarations
\begin{verbatim}
kind  tp              type.
type  form            tp.
type  intty           tp.
type  arrow           tp -> tp -> tp.
type  pair            tp -> tp -> tp.
\end{verbatim}
This change forces several other changes.  The type of \cd{tm} must be
changed to \cd{tp -> ltp}.  The \cd{well\_typed} clause for \cd{tp} is
no longer necessary.  The \cd{istype} predicate and all of the clauses
for it can be removed; all \cd{istype} subgoals in other clauses can
be eliminated.  The \cd{well\_typed} clause for \cd{tm} can also be
eliminated since checking for well-typedness amounts to simply using
the \cd{hastype} predicate.  In the types of all of the constants,
wherever there appears a term \cd{T} of type \cd{ltm} such that \cd{T}
represents an object-logic type, the type of \cd{T} must be changed to
\cd{tp}.  Program~\ref{trans-mod2} illustrates these changes on the
subset of declarations and clauses from Program~\ref{trans-mod1}.
Note that the types of \cd{forall} and \cd{forall\_i} are changed to
reflect the fact that the first argument \cd{T} has type \cd{tp}.
\begin{figure}
\begin{verbatim}
kind  ltp             type.
kind  ltm             type.
kind  tp              type.

type  hastype         ltm -> ltp -> o.
type  proves          ltm -> ltp -> o.
type  well_typed      ltm -> ltp -> o.

type  tm              tp -> ltp.
type  pf              ltm -> ltp.

type  arrow           tp -> tp -> tp.           infixr  arrow   8.
type  forall          tp -> (ltm -> ltm) -> ltm.
type  forall_i        tp -> (ltm -> ltm) -> (ltm -> ltm) -> ltm.

well_typed M (pf A) :- hastype A (tm form), proves M (pf A).

hastype (forall T A) (tm form) :-
  pi x\ (hastype x (tm T) => hastype (A x) (tm form)).

proves (forall_i T A Q) (pf (forall T A)) :-
  pi y\ (hastype y (tm T) => hastype (A y) (tm form)),
  pi y\ (hastype y (tm T) => proves (Q y) (pf (A y))).
\end{verbatim}
\captionx{Program}{Modification of Program~\ref{trans-mod1}.}
\label{trans-mod2}
\end{figure}

Our fourth modification to the \lprolog{} code allows the \lprolog{}
type system to make further useful distinctions for our particular
object logic.
We introduced the \cd{hastype} and \cd{proves} predicate for the cases
when the second argument to our old \cd{hasltype} had the forms
\cd{(tm T)} and form \cd{(pf A)}, respectively.  We can further
simplify these clauses by eliminating the \cd{tm} and \cd{pf}
constants.  Simply eliminating them means we must change the types of
the second argument to these predicates appropriately,
\begin{verbatim}
type  hastype         ltm -> tp -> o.
type  proves          ltm -> ltm -> o.
\end{verbatim}
but we can go a step further than that.  Notice that after removing
\cd{tm} and \cd{pf}, terms appear as the first argument to
\cd{hastype} and types as the second, and that proofs appear as the
first argument to the \cd{proves} predicate and formulas, which are a
subset of the terms, appear as the second.  To make these distinctions
in the program, we reintroduce the constants \cd{tm} and \cd{pf}, but
this time as \lprolog{} types which replace \cd{ltm}.
\begin{verbatim}
kind  tm              type.
kind  pf              type.
type  hastype         tm -> tp -> o.
type  proves          pf -> tm -> o.
\end{verbatim}
After making all the changes discussed so far to the types and clauses
in Programs~\ref{trans-types} and~\ref{trans-rules}, we obtain the
somewhat simpler versions in Programs~\ref{trans-types-a}
and~\ref{trans-rules-a}.
\begin{figure}
\begin{verbatim}
kind  tp              type.
kind  tm              type.
kind  pf              type.

type  hastype         tm -> tp -> o.
type  proves          pf -> tm -> o.
type  well_typed      pf -> tm -> o.

type  form            tp.
type  intty           tp.
type  arrow           tp -> tp -> tp.           infixr  arrow   8.

type  lam             tp -> tp -> (tm -> tm) -> tm.
type  app             tp -> tp -> tm -> tm -> tm.
type  eq              tp -> tm -> tm -> tm.
type  imp             tm -> tm -> tm.           infixr  imp     7.
type  forall          tp -> (tm -> tm) -> tm.
type  false           tm.

type  refl            tp -> tm -> pf.
type  beta            tp -> tp -> (tm -> tm) -> tm -> pf.
type  congr           tp -> tm -> tm -> (tm -> tm) -> pf -> pf -> pf.
type  imp_i           tm -> tm -> (pf -> pf) -> pf.
type  imp_e           tm -> tm -> pf -> pf -> pf.
type  forall_i        tp -> (tm -> tm) -> (tm -> pf) -> pf.
type  forall_e        tp -> (tm -> tm) -> pf -> tm -> pf.
\end{verbatim}
\captionx{Program}{Modified version of Program~\ref{trans-types}.}
\label{trans-types-a}
\end{figure}
\begin{figure}
\begin{verbatim}
well_typed M A :- hastype A form, proves M A.

hastype (lam T1 T2 F) (T1 arrow T2) :-
  pi x\ (hastype x T1 => hastype (F x) T2).
hastype (app T1 T2 F X) T2 :- hastype F (T1 arrow T2), hastype X T1.
hastype (eq T X Y) form :- hastype X T, hastype Y T.
hastype (A imp B) form :- hastype A form, hastype B form.
hastype (forall T A) form :- pi x\ (hastype x T => hastype (A x) form).
hastype false form.

proves (refl T X) (eq T X X) :- hastype X T.
proves (beta T1 T2 F X) (eq T2 (app T1 T2 (lam T1 T2 F) X) (F X)) :-
  pi x\ (hastype x T1 => hastype (F x) T2).
proves (congr T X Z H P1 P2) (H X) :-
  hastype X T, hastype Z T, pi x\ (hastype x T => hastype (H x) form),
  proves P1 (eq T X Z), proves P2 (H Z).
proves (imp_i A B Q) (A imp B) :- hastype A form, hastype B form.
  pi p\ (proves p A => proves (Q p) B).
proves (imp_e A B Q1 Q2) B :- hastype A form, hastype B form,
  proves Q1 (A imp B), proves Q2 A.
proves (forall_i T A Q) (forall T A) :-
  pi y\ (hastype y T => hastype (A y) form).
  pi y\ (hastype y T => proves (Q y) (A y)).
proves (forall_e T A Q Y) (A Y) :-
  pi y\ (hastype y T => hastype (A y) form),
  proves Q (forall T A), hastype Y T.
\end{verbatim}
\captionx{Program}{Modified version of Program~\ref{trans-rules}.}
\label{trans-rules-a}
\end{figure}
Note that \cd{tm} and \cd{pf} no longer appear in clauses
(Program~\ref{trans-rules-a}), and instead appear in types
(Program~\ref{trans-types-a}).  Also note the new type and clause for
\cd{well\_typed} as compared to what they were in
Program~\ref{trans-mod2}.

The types and clauses in Programs~\ref{trans-types-a}
and~\ref{trans-rules-a} are now quite close to those of
Programs~\ref{core-types} and~\ref{core-rules} in Section~\ref{core}.
The remaining changes are optimizations that can be best illustrated
if we view the \lprolog{} code as a proof checker.  In particular, for
any subgoal of the form \cd{(proves P A)}, we assume the proof and the
formula are given at the outset (no logical variables) and that the
subgoal \cd{(hastype A form)} will be asked first (e.g., via the
\cd{well\_typed} predicate).  With this in mind, by looking at some of
the clauses for the \cd{proves} predicate, we find two kinds of
redundancy.  Consider, for example, the clause for \cd{refl}.  The
arguments \cd{T} and \cd{X} appear in both the proof and the formula.
Assuming that a formula and proof are always paired together, any
arguments that appear in the formula do not have to be repeated in the
proof.  Thus we can remove both arguments to \cd{refl}.  Also, since
we assume that the formula has already been typechecked, the
\cd{hastype} subgoal is redundant and can be eliminated.  Thus we
achieve the simple form for the \cd{refl} rule as it appears in
Program~\ref{core-rules}.

Next consider the clause for \cd{imp\_e}.  Since \cd{B} is the formula
whose proof is to be checked, we don't need an extra copy among the
arguments to \cd{imp\_e}.  We also don't need to typecheck \cd{B}
since this has been done via the initial call to \cd{well\_typed}.  If
we are to guarantee correct typing of the formula in \emph{any}
\cd{proves} subgoal generated during proof checking, then we need to
keep \cd{hastype} subgoals for any formula that does not appear as a
subformula of the formula in the head of the clause.  In the
\cd{imp\_e} clause, the goal \cd{(hastype A form)} is asked before
\cd{(proves Q2 A)} and this \cd{hastype} subgoal cannot be removed.
These changes lead to the \cd{imp\_e} clause in
Program~\ref{core-rules}.

Analogously, we can examine the \cd{hastype} clauses and remove
redundant arguments from terms.  For example, in the case of \cd{app},
the type \cd{T2} can be removed because it appears as the second
argument to \cd{hastype}.  We must keep \cd{T1} if we want to preserve
the property that proof checking will not introduce logic variables.

Note that when comparing Program~\ref{trans-rules-a} to
Program~\ref{core-rules}, in the \cd{proves} clause for \cd{congr}, no
arguments are removed from the proof term in either case, even though \cd{H}
and \cd{X} appear in the second argument to \cd{proves}.  The reason is
that backchaining on this clause requires higher-order matching, for
which there can be more than one solution.  One further criteria that
we place on our proof checker is that it cannot backtrack.  Thus we
must include \cd{H} and \cd{X} explicitly in the proof term to prevent
the possibility that when backchaining on this clause, a backtrack
point is created by unification.  We can, however, eliminate the
typechecking subgoal for \cd{H} because its well-typedness follows
from the fact that \cd{(H X)} has type \cd{form} and \cd{X} has type
\cd{T}.  Eliminating this subgoal from the clause in
Program~\ref{trans-rules-a} gives us the clause in
Program~\ref{core-rules}.

After making analogous changes to all of the clauses in
Program~\ref{trans-rules-a}, the only remaining difference in
Program~\ref{core-rules} is the use of \cd{assump} to identify
assumptions added during proof checking,
which as stated earlier, is not necessary, but is useful for various
programming tasks in our proof-carrying code system.

Note that in making changes to the \lprolog{} code, we have been
careful not to complicate proof checking by requiring any more power
from \lprolog{} than was needed to execute the code obtained directly
from the transformation.  The same is not true for the Twelf code.  As
stated earlier, the version that used abbreviations
(Program~\ref{core-twelf}) needs more type reconstruction power than
the version with all arguments explicitly included.

In summary, using the formal correspondence has provided a principled
way to arrive at the versions of the encodings of the object logic in
Twelf (Program~\ref{core-twelf}) and \lprolog\
(Program~\ref{core-rules}) that we have compared.  The main
differences are (1) the Twelf encoding is more concise because
dependent types eliminate the need for explicit typechecking
subgoals, and (2) in \lprolog, unlike Twelf, proof checking of the
optimized version of the encoding is no more complex than proof
checking the original.

\end{document}